\documentclass[review, 1p]{elsarticle}
\usepackage{amsfonts, amsmath, amssymb, amsthm}
\usepackage{tikz}
\usepackage{physics}
\usepackage{siunitx}
\usepackage{wrapfig}
\usepackage{hyperref}
\hypersetup{colorlinks=true,
citecolor=black,
filecolor=black,
linkcolor=black,
urlcolor=black}
\usepackage[version=4]{mhchem}

\DeclareSIUnit\pcm{pcm}
\usetikzlibrary{shapes.geometric}
\usetikzlibrary{arrows.meta}
\usetikzlibrary{calc}
\usetikzlibrary{patterns}
\usepackage{subcaption}

\newcommand{\keff}{k_{\text{eff}}}
\newcommand{\doubleprime}{{\prime\prime}}

\author[1]{Mathis Caprais\corref{cor1}%
\fnref{fn1}}
\ead[url]{mathis.caprais@cea.fr}
\fntext[fn1]{\href{mailto:mathis.caprais@cea.fr}{mathis.caprais@cea.fr}}

\author[2]{André Bergeron}
\ead{andre.bergeron@cea.fr}

\author[1]{Nathan Greiner}
\ead{nathan.greiner@cea.fr}

\author[3]{Daniele Tomatis}
\ead{daniele.tomatis@newcleo.com}

\cortext[cor1]{Corresponding author}
\address[1]{Université Paris-Saclay, CEA, Service d'\'{E}tudes des Réacteurs et de Mathématiques Appliquées, Gif-sur-Yvette, 91191, France}
\address[2]{Université Paris-Saclay, CEA, Service de Thermo-hydraulique et de Mécanique des Fluides, Gif-sur-Yvette, 91191, France}
\address[3]{Newcleo Srl, Via Giuseppe Galliano 27, Torino, 10129, Italy}
\title{A new calculation method using pathlines for delayed neutron precursors in liquid nuclear fuels}
\journal{Nuclear Science and Engineering}
\begin{document}
\begin{frontmatter}
\begin{abstract}
In nuclear reactors, Delayed Neutron Precursors (DNPs) are important for reactor safety and operation. In liquid nuclear fuels, DNPs are transported by the flow, and an advection-reaction balance equation for their concentration must be solved in addition to the Neutron Balance Equation. This research paper applies the method of characteristics to solve the DNPs equation, using techniques previously developed in petroleum engineering and groundwater analysis. The calculation strategy incorporates a pathline generation algorithm covering all the mesh cells of the geometry for concentration calculations, and pathlines are reconstructed for both step and piecewise linear velocity fields. The analytical solutions are used on a Cartesian mesh to compute the DNPs concentrations assuming a step source of DNPs. The results obtained on the steady state phases of a benchmark problem illustrate this new method's capability to solve advection-dominated problems in liquid-fueled reactors. The method developed in this work is suited to smooth velocity fields (e.g., laminar or RANS flows) where pathlines can be tracked, and where DNP diffusivity is negligible.

% The calculation strategy incorporates a pathline generation algorithm that ensures a complete coverage of the geometry, for both piecewise uniform and piecewise linear velocity fields. The MOC for DNPs is tested on the steady state steps of the ``square'' benchmark dedicated to fast-spectrum MSRs from the CNRS. The results highlight the capability of the method to compute the DNPs concentrations in a liquid-fueled reactor. The MOC for DNPs is aimed to be applicable on smooth velocity fields (e.g. laminar or RANS flows) where pathlines are well-defined, and where DNPs diffusivity is first negligible.
\end{abstract}
\end{frontmatter}
\newpageafter{abstract}
\section*{Introduction}
Molten Salt Reactors (MSRs) are nuclear reactors that use molten salts with dissolved fissile material as fuel. These systems were originally developed in the 50s and 60s and prototypes of MSRs were constructed such as the Aircraft Reactor Experiment and the Molten Salt Reactor Experiment \cite{bettis1957aircraft, rosenthal1970molten}. These reactors have been revisited for their potential to provide a sustainable and safe energy source since 2001 in the framework of GenIV International Forum.

The use of liquid nuclear fuels induces a coupling between neutronic and fluid dynamics. Delayed neutrons can be emitted at a position different from the position where the precursors are created due to transport by the velocity field. This phenomenon can contribute to an overall reduction in the reactivity of the system since the precursor's decay occurs far from the region with major power release (core). Existing neutron transport tools must be updated to take into account the transport of DNPs.

Two main approaches have been developed to solve this problem. The first consisting in solving the neutron and DNPs balance equations together within the CFD code as an augmented system of equations, as done in the past with the OpenFOAM platform \cite{Blanco2020, Cervi2019, cervi2019multiphysics}.

The second approach consists of solving the DNPs balance equations separately from the neutron balance equation, within a fluid dynamics code \cite{greiner2023new,griffinCNRS2024}. The DNPs concentrations are then transferred to an existing neutronic code that can solve the neutron flux. This strategy presents the advantage of being able to use existing neutronic codes with little modification. These codes are well-established, validated and have reference solutions for neutron transport.

The balance equation can be analytically inverted to get the precursor concentrations given by an integral equation in the neutron flux, provided that the velocity field is known and that the pathlines exist. The substitution of the solutions for the precursors into the original neutron balance equation yields an integro-differential equation which is of integral type also in space \cite{pazsit2012analytical, Caprais2022}.

The method presented in this work applies the Method Of Characteristics (MOC) to the DNPs balance equation yielding an ODE that has analytic solution in the neutron flux. During a steady state calculation, the neutron source must be updated at each power iteration including all contributions coming from the precursors' decay. We neglect in this work the molecular diffusion of fission products, which would change the nature of the differential equation. The known input velocity field allows calculating a set of pathlines covering all the physical domain. Then, the precursors' concentrations are determined analytically along the pathlines. The calculation of the pathlines and of all their intersections with the problem geometry separating the different materials constitutes the tracking data. The tracking data must be calculated once at the beginning of the calculation, provided a constant geometry in time. This method is applied to the lid-driven benchmark proposed by CNRS for testing multiphysics codes for the simulation of fast-spectrum MSRs \cite{tiberga2020results}. The MOC for DNPs is coupled to a discrete ordinates ($S_{\mathrm{N}}$) solver for the neutron transport equation. The numerical results obtained performing the steady phases of the benchmark are in agreement with the results provided by the other participants to the benchmark exercise. The accuracy of the results depend on the amount of pathlines simulated and on where they are drawn. The starting point of these pathlines is selected randomly, and they are obtained as approximations of the velocity vector assigned to the cells of the mesh upon which the neutron flux equation is solved. The average precursors concentrations in the cells are obtained by proper weighting of the corresponding quantities determined on the pathlines which implies some noise observed in the results with mesh refinement.

The paper is organized as follows. First, the benchmark is presented in Sec. \ref{sec:lid_cavity_pb}. In Sec. \ref{sec:change_of_coordinates}, the DNPs balance equations are introduced, and an analytical solution is derived. In Sec. \ref{sec:calculation_strategy}, the general calculation strategy is presented. Two pathline generation methods based on the level of discretization of the velocity field are presented in Sec. \ref{sec:pathlines_tracking}. Then, in Sec. \ref{sec:discretization}, the integral equation for the precursors is solved by the step approximation of the source. The general precursors' calculation scheme is given in Sec. \ref{sec:general_precursors_calculation_scheme}. The neutron transport equation is recalled and discretized in angle, space and energy in Sec. \ref{sec:coupling}. The results obtained on the steady state phases of the benchmark as well as a performance analysis of the different tracking methods are presented in Sec. \ref{sec:results}. The paper concludes with a discussion on the potential of the MOC of DNPs for the calculation of the delayed neutron source in liquid fuel reactors and its limitations in Sec. \ref{sec:conclusion}.
\section{The lid-driven cavity problem}\label{sec:lid_cavity_pb}
\begin{figure}[h]
\centering
\begin{tikzpicture}[scale=1, every node/.style={scale=1}]
% Draw the square
\draw (0,0) rectangle (4,4);

% Shading for immovable walls
\draw[pattern=north east lines] (0,0) rectangle (-0.1,4);
\draw[pattern=north east lines] (0,0) rectangle (4,-0.1);
\draw[pattern=north east lines] (4,0) rectangle (4.1,4);
%\draw[pattern=north east lines] (0,4) rectangle (4,4.2);

% Dashed lines
\draw[dashed] (0,2) -- (4,2);
\draw[dashed] (2,0) -- (2,4);

% Labels
%   \node at (2,4.5) {B'};
%   \node at (4.5,2) {A'};

% Arrows
\draw[->] (2,4) -- +(1,0) node[above] {\(\vb{u}_0\)};

% Axes
\draw[->] (-0.5,-0.5) -- (-0.5,0.5) node[left] {$\vu{y}$};
\draw[->] (-0.5,-0.5) -- (0.5,-0.5) node[below] {$\vu{x}$};

% Points A and B
\fill (0,2) circle (1pt) node[below right] {A};
\fill (2,4) circle (1pt) node[below left] {B};
\fill (2,0) circle (1pt) node[above left] {B'};
\fill (4,2) circle (1pt) node[below left] {A'};

\end{tikzpicture}
\caption{Sketch of the problem considered in this work.}
\label{fig:lid_cavity}
\end{figure}
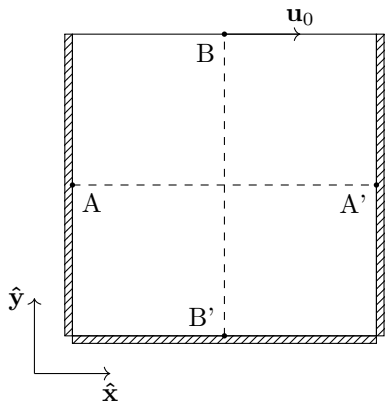

The lid-driven cavity problem is a standard laminar 2D benchmark problem in computational fluid dynamics \cite{burggraf1966analytical, ferziger2019computational}. It consists of a square cavity of \SI{2}{\meter} by \SI{2}{\meter} with a moving lid, as depicted in Fig. \ref{fig:lid_cavity}. This problem exhibits viscous eddies and recirculation zones where pathlines form closed and open tracks.
This computational fluid dynamics problem has been adapted and used in previous works to benchmark codes for the simulation of liquid-fuel nuclear reactors \cite{tiberga2020results, greiner2023new, groth2021verification}. The original benchmark is modified to simulate a homogeneous liquid-fueled reactor. All the specifications of the benchmark are given in a previous reference \cite{tiberga2020results}. The geometry is discretized into a Cartesian mesh of $100\times100$ cells. The mesh is equidistant, and the cells are squares. The velocity field is incompressible and the flow is considered time-independent.
\section{Pathlines change of coordinates}\label{sec:change_of_coordinates}
This section covers a change of coordinates that transforms the advection balance partial differential equation into an ordinary differential equation on pathlines. In Sec. \ref{subsec:precursors_balance}, the general balance equation for DNPs is reminded. This equation is scaled in Sec. \ref{subsubsec:nondimensionalization} to highlight its dominant terms. For the laminar problem considered in this work, advection dominates molecular diffusion. Then, pathlines coordinates are defined in Sec. \ref{subsec:pathlines_coordinates} and the DNPs balance equations as well as their analytical solutions are presented in Sec. \ref{subsec:change_of_coordinates}. 
\subsection{Precursors balance equation}\label{subsec:precursors_balance}
DNPs are advected, and potentially diffused in space. The balance equation for the precursor concentration $C_j$ (\SI{}{\per\cubic\meter}) for the $j$-family for an incompressible liquid fuel in the steady state is,
\begin{equation}
    \vb{u}\cdot\grad{C_j} - \div{D\grad{C_j}}+ \lambda_j C_j = S_j,
    \label{eq:precursors_advection_diffusion}
\end{equation}
with $\vb{u}$ (\SI{}{\meter\per\second}) being the velocity vector field, $D$ (\SI{}{\square\meter\per\second}) is the diffusivity and $\lambda_j$ (\SI{}{\per\second}) the decay constant of the $j$-family. $S_j = \beta_j R_f$ is the source term with $\beta_j$ (dimensionless) being the delayed neutron fraction and $R_f$ (\SI{}{\per\cubic\meter\per\second}) the total $\nu$-fission reaction rate which is given in Sec. \ref{sec:coupling} by Eq. \eqref{eq:delayed_fission_source}.
\subsubsection{Dimensional Analysis}\label{subsubsec:nondimensionalization}
The relative prominence of the different physical phenomena accounted for in Eq. \eqref{eq:precursors_advection_diffusion} can be assessed by scaling the equation to obtain dimensionless quantities. The scaled quantities are:
\begin{equation*}
    \vb{u}^\prime = \frac{\vb{u}}{u_0}, \quad \vb{r}^\prime = \frac{\vb{r}}{L}, \quad C_j^\prime = \frac{C_j}{C_0}, \quad S_j^\prime = \frac{S_j}{S_0}, \quad D^\prime = \frac{D}{D_0},
\end{equation*}
where $u_0$, $L$, $C_0$, $S_0$ and $D_0$ are the characteristic velocity, length, concentration, source and diffusivity, respectively. The DNPs source and the DNPs concentration are not independent quantities. They are proportional because the balance equation Eq. \eqref{eq:precursors_advection_diffusion} is linear, $S_0 \propto \lambda_j C_0$. The scaled balance equation thus reads:
\begin{equation}
    \vb{u}^\prime\cdot\grad{C_j^\prime} - \frac{1}{\mathcal{A}}\div{D^\prime\grad{C_j^\prime}}+  \frac{1}{\mathcal{B}}C_j^\prime = \frac{S_j^\prime}{\mathcal{B}},
    \label{eq:precursors_advection_adim}
\end{equation}
with characteristic parameters defined as,
\begin{equation}
    \mathcal{A} = \frac{L u_0}{D_0} \qq{and} \mathcal{B} = \frac{u_0}{L \lambda_j},
    \label{eq:adim_parameters}
\end{equation}
representing respectively the ratio of advection over diffusion and the ratio of advection over precursor decay. The first dimensionless number identifies as a Péclet number for mass transport. The second dimensionless number identifies as an advection-reaction number, which can be seen as a variation of the first Damköhler number \cite{weiland2020mechanics}. In a molten salt, the diffusion coefficient can be estimated as $D = \nu/\text{Sc}$, with $\nu$ being the kinematic viscosity and $\text{Sc}$ the Schimdt number \cite{di2022multiphysics}. For the salt considered, the Schimdt number is equal to \SI{2e8}{}, the kinematic viscosity is equal to \SI{2.5e-2}{\meter\squared\per\second} and the diffusion coefficient is equal to \SI{1.25e-10}{\meter\squared\per\second} \cite{di2022multiphysics}. The characteristic length is the size of the cavity, $L=\SI{2}{\meter}$. The characteristic velocity is \SI{5e-1}{\meter\per\second}. The average decay constant of the DNPs is equal to \SI{1e-1}{\per\second}. With this knowledge, the characteristic parameters of Eq. \eqref{eq:adim_parameters} can be estimated,
\begin{equation*}
    \mathcal{A} = \SI{8e9}{} \qq{and} \mathcal{B} = \SI{2.5}{}.
\end{equation*}
Both $\mathcal{A}$ and $\mathcal{B}$ are compared to one, which is the scale of the advection term in Eq. \eqref{eq:precursors_advection_adim}. Since $\mathcal{A}\ggg 1$, advection dominates diffusion and Eq. \eqref{eq:precursors_advection_diffusion} reduces to,
\begin{equation}
    \vb{u}\cdot\grad{C_j} + \lambda_j C_j = S_j.
    \label{eq:precursors_advection}
\end{equation}
Considering Eq. \eqref{eq:adim_parameters}, it should be emphasized that the characteristic parameter $\mathcal{A}$ depends on the decay constant of the DNPs. Since decay constants range from \SI{1e-2}{\per\second} to \SI{1e0}{\per\second} \cite{tiberga2020results}, short-lived precursors are less affected by transport than long-lived precursors.
\subsection{Pathlines coordinates}\label{subsec:pathlines_coordinates}
Pathlines are the trajectories of the fluid particles and are defined by the kinematic equation,
\begin{equation}
    \dv{\vb{r}}{\tau} = \vb{u}\qty(\vb{r}) \qq{with} \vb{r}(0) = \vb{r}_0,
    \label{eq:pathlines}
\end{equation}
where $\tau$ is called the time-of-flight and is the time spent by a particle along the trajectory and $\vb{r}_0$ is the initial position of the fluid particle. These particles are considered \textit{virtual} or \textit{passive} as they do not interact back with the flow. Due to the incompressible nature of the fluid flow, 
\begin{equation}
    \div{\vb{u}} = 0,
    \label{eq:incompressible}
\end{equation}
a potential vector of the velocity field $\vb{u}$ can be defined and represented as,
\begin{equation}
    \vb{u} = \curl{\vb*{\Pi}} \qq{and} \vb*{\Pi} = \grad{\varphi} + \chi\grad{\psi},
    \label{eq:potential}
\end{equation}
where $\varphi$, $\chi$ and $\psi$ are scalar fields and usually called Monge or Clebsch potentials. This representation of the potential vector yields to a new representation of the velocity field. Using Eq. \eqref{eq:potential}, the velocity field can be written as,
\begin{equation}
    \vb{u} = \grad{\chi} \crossproduct \grad{\psi},
    \label{eq:velocity_representation}
\end{equation}
which is the representation of the velocity field in terms of potentials. The triplet $\qty(\tau, \chi, \psi)$ is a new set of coordinates that can be used to describe the flow. Along a pathline, both $\chi$ and $\psi$ are constant as their gradients are orthogonal to the velocity field $\qty(\vb{u}\cdot\grad{\chi} = \vb{u}\cdot\grad{\psi}=0)$ due to the representation of Eq. \eqref{eq:velocity_representation} being a cross-product. The potentials introduced in the velocity field are known as the bistream function and induces a change of coordinate that conserves the infinitesimal volume element $\dd{V} = \dd{x}\dd{y}\dd{z} = \dd{\tau}\dd{\chi}\dd{\psi}$, which is a consequence of the incompressibility of the flow, Eq. \eqref{eq:incompressible}. In the scope of this work which is the lid-driven cavity problem, the bistream function identifies as the streamfunction \cite{ferziger2019computational}.
\subsection{Change of coordinates}\label{subsec:change_of_coordinates}
The advection balance equation, Eq. \eqref{eq:precursors_advection}, can be rewritten along a pathline using the chain rule,
\begin{equation*}
    \dv{}{\tau}C_j = \sum_i \pdv{x_i}{\tau}\pdv{C_j}{x_i} = \vb{u}\cdot \grad{C_j}.
    \label{eq:precursors_advection_pathlines}
\end{equation*}
effectively transforming the partial differential equation into an ordinary differential equation,
\begin{equation*}
    \dv{}{\tau}C_j\qty(\vb{r}\qty(\tau))  + \lambda_j C_j\qty(\vb{r}\qty(\tau))  = S_j\qty(\vb{r}\qty(\tau)) .
    \label{eq:precursors_advection_pathlines_ode}
\end{equation*}
The analytical solution of this first order differential equation is an integral equation,
\begin{equation}
    C_j \qty(\vb{r}\qty(\tau)) 
    = C_j\qty(\vb{r}\qty(\tau_{0}))e^{-\lambda_j \qty(\tau -\tau_{0})}
    + \int_{\tau_{0}}^{\tau}\dd{\tau^\prime} S_j \qty(\vb{r}\qty(\tau^\prime))e^{-\lambda_j \qty(\tau - \tau^\prime)} \qq{with} \tau_{0} < \tau.
    \label{eq:balance_precs_sol_incomp}
\end{equation}
Eq. \eqref{eq:balance_precs_sol_incomp} integrates the DNPs source term along the pathline multiplied by an exponentially-decaying term and assumes that the concentration of DNPs is known at some point or some time-of-flight $\tau_0$.
\subsection{The special case of closed pathlines}
If the pathline forms a closed circuit where the continuity of the solution along the line. Given a recirculation period $\tau_l$; the boundary condition is $C_j\qty(\vb{r}\qty(\tau_l)) = C_j\qty(\vb{r}\qty(0))$. The concentration along a closed pathline is given by,
\begin{equation}
    C_j \qty(\vb{r}(\tau)) = \frac{e^{-\lambda \tau_l}}{1 - e^{-\lambda \tau_l}} \int_0^{\tau_l}\dd{\epsilon}S_j\qty(\vb{r}(\epsilon))e^{-\lambda_j \qty(\tau - \epsilon)} + \int_{0}^{\tau}\dd{\epsilon} S_j \qty(\vb{r}\qty(\epsilon))e^{-\lambda_j \qty(\tau - \epsilon)},
    \label{eq:transmission_equation_closed}
\end{equation}
where the first integral accounts for the contribution of the previous recirculations along the closed trajectory. The factor in front of the first integral can be expanded as,
\begin{equation*}
    \frac{e^{-\lambda \tau_l}}{1 - e^{-\lambda \tau_l}} = \sum_{n=1}^{\infty} e^{-n\lambda \tau_l},
\end{equation*}
which accounts for an infinite recirculation on a closed trajectory. This development in series was already given in analytical studies of the 1D molten salt reactor \cite{pazsit2012analytical,Caprais2022}.
\section{A new calculation method for delayed neutron precursors}\label{sec:calculation_strategy}
The calculation method for the DNPs concentration uses the analytical solutions of Sec. \ref{subsec:change_of_coordinates}. The method is based on the reconstruction of pathlines on the velocity field. A tracking strategy covering the whole mesh is presented in Sec. \ref{subsec:general_tracking_strategy}. The pathline reconstruction within a cell is presented in Sec. \ref{sec:pathlines_tracking}, and the approximation of the integral equation is presented in Sec. \ref{sec:discretization}. The scheme for the calculation of the precursor concentrations is presented in Sec. \ref{sec:general_precursors_calculation_scheme}. 
\subsection{General tracking strategy}\label{subsec:general_tracking_strategy}
To evaluate the concentrations, an algorithm generating pathlines that cross every cell of the mesh is necessary to avoid cells without information about precursors. The algorithm is presented in Sec. \ref{subsec:pathlines_generation}. The tracking algorithm places seeds (starting point of a pathline with $\tau=0$) at the center of the cells and reconstructs them forward and backward in time. Reverse time integration is achieved by exploiting the odd time-of-flight symmetry of the kinematic equation Eq. \eqref{eq:pathlines}, $\tau \to -\tau$ gives $\vb{u} \to -\vb{u}$.
Both closed and open pathlines are used to cover the mesh. We consider that a pathline closes upon itself if it enters and leaves the same cell during the reconstruction process.
\subsubsection{Pathlines generation}\label{subsec:pathlines_generation}
\begin{figure}[h!]
\centering
\includegraphics[scale=0.7]{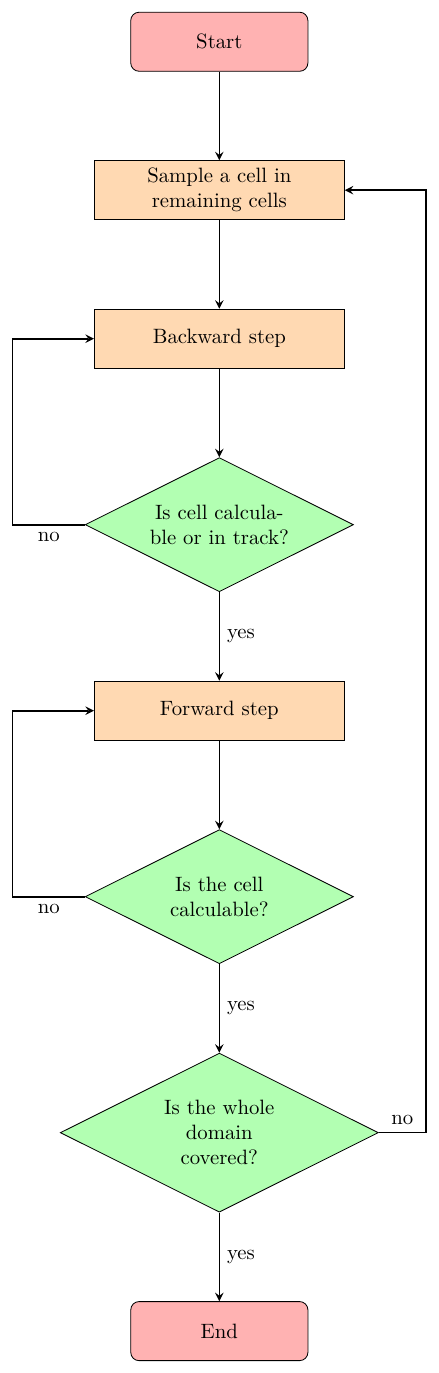}
\caption{Flowchart of the seeding strategy.}
\label{fig:calculation_strategy}
\end{figure}

The generation strategy is presented on the flowchart in Fig. \ref{fig:calculation_strategy} with the goal of generating pathlines that cross every cell of the mesh.
The DNPs solution is known analytically along both open and closed pathlines, with different initial conditions though. Open pathlines use the entering concentration values at the boundary of the calculation domain, while continuity applies on closed boundaries as shown in section X. The computability status of the cells are stored in a computability table. Thus, the algorithm goes as follows. New pathlines are determined every time that uncrossed cells are found. Specifically, a new pathline starts from the center of an uncrossed cell that is randomly selected. A cell is randomly sampled where the computability table is \textit{False}, and a seed point is placed in this cell. A pathline is integrated backward in time from this point until reaching a computable cell, or reaching a cell that has already been crossed, thereby forming a closed-loop pathline. This first phase marks a finite number of cells as computable. Then, starting from the same seed point, the pathline is integrated forward in time until forming a closed-loop pathline or reaching a calculable computable cell. This process defines two computable tracks. The algorithm then moves to the next cell where the computability table is \textit{False} and repeats the process. The algorithm stops when all the cells are marked as computable.
\subsection{Pathlines tracking}\label{sec:pathlines_tracking}
The integration of the pathline equation Eq. \eqref{eq:pathlines} relies on the velocity field provided by the CFD code, TrioCFD. Two types of velocity fields are considered: an averaged velocity vector within each cell and velocity vector components defined on the cell faces. The former is a limiting case of the latter when the velocity is uniform within the cell.
\subsubsection{First-order interpolation}\label{subsec:first_order}
This pathline integration assumes that the vector components are known at the center of each face, the velocity components can be assumed to evolve linearly within the cell.
\begin{figure}[h]
\centering
\begin{tikzpicture}[>=Stealth]
\coordinate (g) at (0,0);
\coordinate (g1) at (0,4);
\coordinate (g2) at (4,4);
\coordinate (g3) at (4,0);

\draw (g) -- (g1) node[midway, above, sloped, xshift=-0.5cm, yshift=0.0cm] {West};
\draw (g1) -- (g2) node[midway, above, sloped, xshift=0.5cm, yshift=0.0cm] {North};
\draw (g2) -- (g3) node[midway, above, sloped, xshift=-0.5cm, yshift=0.0cm] {East};
\draw (g3) -- (g) node[midway, below, sloped, xshift=-0.5cm, yshift=-0.0cm] {South};

\draw[->] ($(g1)!0.5!(g)$)++(-1, 0) -- +(2,0) node[above right] {$u_w$};

\draw[->] ($(g1)!0.5!(g2)$)++(0, -1) -- +(0,2) node[above right] {$u_n$};

\draw[->] ($(g2)!0.5!(g3)$)++(-1,0) -- +(2,0) node[above right] {$u_e$};

\draw[->] ($(g3)!0.5!(g)$)++(0, -1) -- +(0,2) node[above right] {$u_s$};

\draw[<->] ([yshift=1.5cm]g1) -- ([yshift=1.5cm]g2) node[midway, above, sloped] {$\Delta x$};
\draw[<->] ([xshift=1.5cm]g2) -- ([xshift=1.5cm]g3) node[midway, above, sloped] {$\Delta y$};

\coordinate (ref_point) at (-1, -1);
\draw[->] (ref_point) -- +(1,0) node[below] {$\vu{x}$};
\draw[->] (ref_point) -- +(0,1) node[left] {$\vu{y}$};

\node[below] at (g) {$\vb{r}_\star$};
\node[circle, fill, inner sep=1pt] at (g) {};

\coordinate (l1) at (2, 2);
\node[above] at (l1) {$\vb{r}_0$};
\node[circle, fill, inner sep=1pt] at (l1) {};

\coordinate (l2) at (3.5, 4);
\node[above right] at (l2) {$\vb{r}_e$};
\node[circle, fill, inner sep=1pt] at (l2) {};
\draw (l1) to[out=0,in=-90] (l2);

\coordinate (l3) at (0.3, 0);
\draw[dashed] (l1) to[out=180,in=90] (l3);

\end{tikzpicture}
\caption{A cell with velocity components known on the faces and a forward (backward) hyperbolic (dashed) trajectory.}
\label{fig:first_order_velocity}
\end{figure}
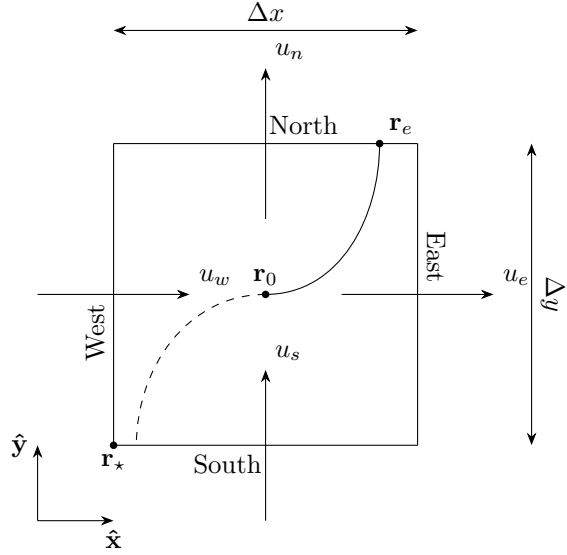
In Fig. \ref{fig:first_order_velocity}, the velocity is defined as,
\begin{equation}
u_x = u_w + \frac{u_e - u_w}{\Delta x}(x-x_\star) = u_w + \alpha_x (x-x_\star),
\label{eq:first_order_velocity_x}
\end{equation}
and,
\begin{equation}
u_y = u_s + \frac{u_n - u_s}{\Delta y}(y-y_\star) = u_s + \alpha_y (y-y_\star).
\label{eq:first_order_velocity_y}
\end{equation}
This is known as Pollock's tracking method and is used in petroleum engineering for streamline simulation \cite{pollock1988semianalytical,datta2007stream}. The pathline trajectory within the cell can be obtained analytically by integrating Eq. \eqref{eq:pathlines}. From a starting point $\vb{r}_0$ within the cell (or on a boundary) the trajectory is,
\begin{equation}
\vb{r}\qty(\tau) = \vb{r}_0
+ \begin{pmatrix}
    \frac{\qty(u_w + \alpha_x (x_0 - x_\star))\qty(\exp(\alpha_x \tau) - 1)}{\alpha_x} \\
    \frac{\qty(u_s + \alpha_y (y_0 - y_\star))\qty(\exp(\alpha_y \tau) - 1)}{\alpha_y}
\end{pmatrix}
\label{eq:P1_trajectory}
\end{equation}
The exit face is determined by selecting the minimal time-of-flight to reach a boundary. For $x$-boundaries,
\begin{equation}
\tau_{w} = \frac{1}{\alpha_x}\ln(\frac{u_w}{u_w + \alpha_x (x_0 - x_\star)}) \qq{and} \tau_e = \frac{1}{\alpha_x}\ln(\frac{u_e}{u_w + \alpha_x (x_0 - x_\star)}).
\label{eq:first_order_tof_x}
\end{equation}
The time-of-flights to reach a $y$-boundary are,
\begin{equation}
\tau_s = \frac{1}{\alpha_y}\ln(\frac{u_s}{u_s + \alpha_y (y_0 - y_\star)}) \qq{and} \tau_n = \frac{1}{\alpha_y}\ln(\frac{u_n}{u_s + \alpha_y (y_0 - y_\star)}).
\label{eq:first_order_tof_y}
\end{equation}
Eqs. \eqref{eq:first_order_tof_x} and \eqref{eq:first_order_tof_y} follow from a straightforward integration of Eq. \eqref{eq:pathlines} with Eqs. \eqref{eq:first_order_velocity_x} and \eqref{eq:first_order_velocity_y}. If the fuel velocity does not change along the $x$ or $y$ axis, then $\alpha_x \to 0$ or $\alpha_y \to 0$. In this special case, the equation of trajectory Eq. \eqref{eq:P1_trajectory} along the axis where the velocity is constant becomes a straight line. If the velocity does not vary along both axes, then this reverts to the case where the velocity vector is constant per cell. Along the $x$-axis, the trajectory is,
\begin{equation*}
x\qty(\tau) \simeq x_0 + \frac{\qty(u_w + \alpha_x x_0)\alpha_x \tau}{\alpha_x} = x_0 + u_w \tau,
\end{equation*}
and the same goes for the $y$-axis when $\alpha_y \to 0$. The time-of-flights become,
\begin{equation*}
\quad \tau_{w} = \frac{x_\star - x_0}{u_x}, \quad\tau_{e} = \frac{x_\star + \Delta x - x_0}{u_x}, \quad \tau_{s} = \frac{y_\star - y_0}{u_y} \qq{and} \tau_{n} = \frac{y_\star + \Delta y - y_0}{u_y}.
\label{eq:time_of_flight_x_oriented_P0}
\end{equation*}
% and to reach an $x$-oriented face,
% \begin{equation*}
% \tau_{s} = \frac{y_\star - y_0}{u_y} \qq{and} \tau_{n} = \frac{y_\star + \Delta y - y_0}{u_y}.
% \label{eq:time_of_flight_y_oriented_P0}
% \end{equation*}
In the remainder of this work, this first-order tracking method is referred to as $\mathcal{P}_1$, while if the velocity is constant within the cell, the method is referred to as $\mathcal{P}_0$.
\subsection{Discretization}\label{sec:discretization}
This subsection covers the numerical method used to solve the balance equation of DNPs along pathlines. The discretization aims to evaluate the volume-average concentration within a cell based on the analytical solution of the balance equation along the pathlines from Eq. \eqref{eq:balance_precs_sol_incomp}. The numerical treatment covers both open and closed pathlines.
\subsubsection{Discretized transmission equation}\label{subsec:discretized_transmission}
The calculation domain is partitioned into a set of cells $D = \bigcup_{k}D_k$ with $k$ being the cell index. The step-source approximation is made for the DNPs source term appearing in the integral term of Eq. \eqref{eq:balance_precs_sol_incomp}, yielding:
\begin{equation}
C^{(k)}_{j,i}\qty(\tau_{i}^{(k),\doubleprime}) - C^{(k)}_{j,i}\qty(\tau_{i}^{(k),\prime}) = \qty(\frac{S_{j}^{(k)}}{\lambda_{j}} - C^{(k)}_{j,i}\qty(\tau_{i}^{(k),\prime}))\qty(1-e^{-\lambda_{j}T_{i}^{(k)}}).
\label{eq:trans_dif}
\end{equation}
with $T_{i}^{(k)} = \tau_{i}^{(k),\doubleprime} - \tau_{i}^{(k),\prime}$, $\tau_{i}^{(k),\doubleprime}$ being the exit time-of-flight from the cell $k$ and $\tau_{i}^{(k),\prime}$ the time-of-flight at the entry of the cell $k$, all on the $i$-pathline. The concentration lying on open pathlines must be known at the entry point of every cell in the mesh up to the boundary of the domain, where boundary  conditions for the entering velocity field apply.
\paragraph{Initial concentration on a closed pathline}
The calculation of the concentration on closed pathlines needs different contributions cause by repeated circulation as shown in Eq. \eqref{eq:transmission_equation_closed}. After choosing a starting point on the pathline, the contribution due to the previous circulation is:
\begin{equation}
    C^i_j\qty(\vb{r}\qty(0)) = \frac{1}{1 - e^{-\lambda_j \tau_{l,i}}} \sum_{k^\prime} \frac{S_j^{(k^\prime)}}{\lambda_j}e^{-\lambda_j \qty(\tau_{l,i} - \tau_i^{(k^\prime),\doubleprime})}\qty(1 - e^{-\lambda_j T_i^{(k^\prime)}}).
    \label{eq:initial_condition_closed}
\end{equation}
In Eq. \eqref{eq:initial_condition_closed}, $k^\prime$ refers to the index of the cells crossed by the closed trajectory. The recirculation time is given by the sum of all the time-of-flights along the loop. If the seed is not placed on a cell boundary, the time-of-flight of the first cell is calculated by integrating the seed forward and backward within the cell until the trajectory reaches a cell boundary. The time-of-flight of the cell is then given by the sum of the forward and backward time-of-flights.
\subsubsection{Discretized line-averaged concentration}\label{subsec:discretized_line_averaged}
The evaluation of the volume-average concentration within a cell requires the calculation of the mean concentration over a pathline within the cell,
\begin{equation}
    \expval{C^{(k)}_{j,i}} = \frac{1}{T_{i}^{(k)}}\int_{\tau_{i}^{(k),\prime}}^{\tau_{i}^{(k),\doubleprime}}\dd{\epsilon}C_{j,i} \qty(\vb{r}\qty(\epsilon)) = \frac{S_{j}^{(k)}}{\lambda_{j}} - \frac{\Delta^{(k)}_{j,i}}{T_{i}^{(k)} \lambda_{j}},
    \label{eq:mean_line}
\end{equation}
where $\Delta^{(k)}_{j,i}=C^{(k)}_{j,i}\qty(\tau_{i}^{(k),\doubleprime}) - C^{(k)}_{j,i}\qty(\tau_{i}^{(k),\prime})$ is given by the discretized equation Eq. \eqref{eq:trans_dif}. If the liquid fuel becomes static, the line-averaged concentration reduces to the static DNPs concentration as Eq. \eqref{eq:mean_line} tends towards $S_{j}^{(k)}/\lambda_{j}$ when $T_{i}^{(k)} \to \infty$.
\subsubsection{Volume-average concentration}\label{subsec:volume_average_concentration}
The volume-averaged concentration within the $k$-cell is then given by a weighted sum of the averaged concentration over the tracks within the cell,
\begin{equation}
    C_j^{(k)} = \frac{1}{V_k}\int_{V_k} \dd{V} C_j\qty(\vb{r}) \simeq \frac{\sum_{i=1}^N T_{i}^{(k)} q_{i}^{(k)} \expval{C_{j,i}^{(k)}}}{\sum_{i=1}^N T_{i}^{(k)} q_{i}^{(k)}} = \sum_{i=1}^N \omega_{i}^{(k)}\expval{C_{j,i}^{(k)}},
    \label{eq:vol_average}
\end{equation}
where $N$ pathlines cross the cell $k$ and $q_i$ is the volume flow rate associated to the $i$-pathline. The weight $\omega_{i}^{(k)}$ corresponding to the pathline $i$ crossing the volume $k$ is obtained by changing the variables under the integral to $\qty(\tau, \chi, \psi)$ and breaking the integral into a discrete sum of pathlines intercepting the volume,
\begin{equation*}
\int_{V_k}\dd{x}\dd{y}\dd{z}=\int_{V_k} \dd{\tau}\dd{\chi}\dd{\psi} \simeq \sum_{i\cap V_k} \int_{\tau_{i}^{(k),\prime}}^{\tau_{i}^{(k),\doubleprime}} \dd{\tau}\int_{S_{i}^{(k)}} \dd{\chi}\dd{\psi}.
\label{eq:volume_flow_rate}
\end{equation*}
The integral over $\chi$ and $\psi$ is identified to be the volume flow rate of the pathline $i$, which serves to define $q_{ik}$ as transversal flow area along the pathline $i$ in cell $k$. $S_{i}^{(k)}$ is delimited by the box bounding $\chi \in [\chi_1, \chi_2] \otimes \psi \in [\psi_1, \psi_2]$.
The coefficients $q$'s arising in the numerical quadrature in space correspond to the average volume flow rate through the transversal surfaces, even if the flow speed does not appear explicitly in the original integral. This follows from the coordinate change from Cartesian to bistream function and time-of-flight.
The volume flow rate associated to the $i$-th pathline is calculated at the beginning of the trajectory as,
\begin{equation*}
q_i = \int_{S_{i}^{(k)}} \dd{S}\vb{n} \vdot \vb{u} = \int_{\psi_1}^{\psi_2}\dd{\psi} \int_{\chi_1}^{\chi_2}\dd{\chi} = \qty(\psi_2 - \psi_1)\qty(\chi_1 - \chi_2)\simeq \norm{\vb{u}}\delta A_i,
\end{equation*}
where $\delta A_i$ is a small surface element associated to the $i$-th pathline. As two constant values of both $\chi$ and $\psi$ (or two constant values of $\chi$ in 2D) define a streamtube, the volume flow rate is constant along a trajectory thanks to mass conservation. Therefore, the explicit tracking of the transversal cross-section or the calculation of the Clebsch potentials, Eq. \eqref{eq:potential}, is not needed.
At equal line-averaged concentrations, the longer the particle stays in the cell, the higher is its contribution to the mean concentration calculated. Volume preservation requires that the sum of all the weights associated to the trajectories is equal to one,
\begin{equation}
\sum_{i=1}^N \omega_{i,k} = 1.
\label{eq:normalization}
\end{equation}
Eq. \eqref{eq:normalization} can be used as normalization constraint for the weights when they are determined by numerical approximations.
\subsection{General precursors' calculation scheme}\label{sec:general_precursors_calculation_scheme}
The first step in the DNPs calculation involves generating a set of pathlines that cover the entire mesh, ensuring that each cell is crossed at least once; this process is referred to as ``tracking.'' This single set of pathlines can be used to compute the DNPs concentrations for all families. From this point, two methods are tested for computing the concentration within a cell: the naive tracking stacking method and the submesh method. The naive tracking stacking method is discussed in Sec. \ref{subsec:naive_tracking}, while the submesh method is detailed in Sec. \ref{subsec:submesh}.
\subsubsection{Naive tracking stacking}\label{subsec:naive_tracking}
To evaluate the concentration within a cell, the naive approach consists of stacking trackings, each providing a different line-averaged concentration within the cell. The concentration within the cell is then computed using Eq. \eqref{eq:vol_average}, assuming equal volume flow rates associated with each pathline. Thus, the weights are assumed to be \(T_i^{(k)} / \sum_j T_j^{(k)}\) for the $k$-th cell.
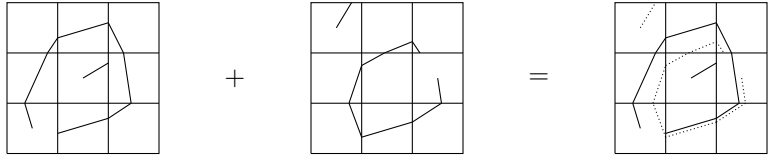
\begin{figure}[h]
\centering
\begin{tikzpicture}
    \coordinate (g) at (0,0);
    \coordinate (g1) at (0,2);
    \coordinate (g2) at (2,2);
    \coordinate (g3) at (2,0);
    
    \draw (g) -- (g1) -- (g2) -- (g3) -- (g);
    
    \coordinate (l1) at ({0, 2/3});
    \coordinate (l2) at ({0, 4/3});
    
    \coordinate (l3) at ({2/3}, 2);
    \coordinate (l4) at ({4/3}, 2);
    
    \coordinate (l5) at (2, {2/3});
    \coordinate (l6) at (2, {4/3});
    
    \coordinate (l7) at ({2/3}, 0);
    \coordinate (l8) at ({4/3}, 0);
    
    \draw (l1) -- (l5);
    \draw (l2) -- (l6);
    \draw (l3) -- (l7);
    \draw (l4) -- (l8);
    
    \coordinate (p1) at ({2/6, 2/6});
    \coordinate (p2) at ({2/6 - 0.1, 2/3});
    \draw (p1) -- (p2);
    
    \coordinate (p3) at ({2/6 + 0.2, 4/3});
    \draw (p2) -- (p3);
    
    \coordinate (p4) at ({2/3, 4/3 + 0.2});
    \draw (p3) -- (p4);
    
    \coordinate (p5) at ({4/3, 4/3 + 0.4});
    \draw (p4) -- (p5);
    
    \coordinate (p6) at ({4/3 + 0.2, 4/3});
    \draw (p5) -- (p6);
    
    \coordinate (p7) at ({4/3 + 0.3, 2/3});
    \draw (p6) -- (p7);
    
    \coordinate (p8) at ({4/3, 2/3 - 0.2});
    \draw (p7) -- (p8);
    
    \coordinate (p9) at ({2/3, 2/3 - 0.4});
    \draw (p8) -- (p9);
    
    \coordinate (p10) at ({1, 1});
    \coordinate (p11) at ({4/3, 1.2});
    \draw (p10) -- (p11);
    
    \node at (3, 1) {$+$};
    
    \coordinate (g4) at (4,0);
    \coordinate (g5) at (4,2);
    \coordinate (g6) at (6,2);
    \coordinate (g7) at (6,0);
    
    \draw (g4) -- (g5) -- (g6) -- (g7) -- (g4);
    \coordinate (k9) at ({4, 2/3});
    \coordinate (k10) at ({4, 4/3});
    
    \coordinate (k11) at ({2/3 + 4}, 2);
    \coordinate (k12) at ({4/3 + 4}, 2);
    
    \coordinate (k13) at (6, {2/3});
    \coordinate (k14) at (6, {4/3});
    
    \coordinate (k15) at ({2/3 + 4}, 0);
    \coordinate (k16) at ({4/3 + 4}, 0);
    
    \draw (k11) -- (k15);
    \draw (k12) -- (k16);
    \draw (k9) -- (k13);
    \draw (k10) -- (k14);
    
    \coordinate (c1) at ({6 - 2/6, 1});
    \coordinate (c2) at ({6 - 2/6 + 0.05, 2/3});
    \node at (7, 1) {$=$};
    \draw (c1) -- (c2);
    
    \coordinate (c3) at ({4 + 4/3, 2/3 - 0.2 - 0.05});
    \draw (c2) -- (c3);
    
    \coordinate (c4) at ({4 + 2/3, 2/3 - 0.2 - 0.05 - 0.2});
    \draw (c3) -- (c4);
    
    \coordinate (c5) at ({4 + 2/3 - 1/6, 2/3});
    \draw (c4) -- (c5);
    
    \coordinate (c6) at ({4+2/3, 2/3 + 0.5});
    \draw (c5) -- (c6);
    
    \coordinate (c7) at ({4+2/3 + 0.3, 4/3});
    \draw (c6) -- (c7);
    
    \coordinate (c8) at ({4+4/3, 4/3 + 0.15});
    \draw (c7) -- (c8);
    
    \coordinate (c9) at ({4+4/3 +0.1, 4/3});
    \draw (c8) -- (c9);
    
    \coordinate (c10) at ({4+2/6, 5/3});
    \coordinate (c11) at ({4+2/6 + 0.2, 2});
    \draw (c10) -- (c11);
    
    \coordinate (a4) at (8,0);
    \coordinate (a5) at (8,2);
    \coordinate (a6) at (10,2);
    \coordinate (a7) at (10,0);
    
    \draw (a4) -- (a5) -- (a6) -- (a7) -- (a4);
    \coordinate (l9) at ({4+4, 2/3});
    \coordinate (l10) at ({4+4, 4/3});
    
    \coordinate (l11) at ({2/3 + 4+4}, 2);
    \coordinate (l12) at ({4/3 + 4+4}, 2);
    
    \coordinate (l13) at (6+4, {2/3});
    \coordinate (l14) at (6+4, {4/3});
    
    \coordinate (l15) at ({2/3 + 4+4}, 0);
    \coordinate (l16) at ({4/3 + 4+4}, 0);
    
    \draw (l11) -- (l15);
    \draw (l12) -- (l16);
    \draw (l9) -- (l13);
    \draw (l10) -- (l14);
    
    \coordinate (d1) at ({8+2/6, 2/6});
    \coordinate (d2) at ({8+2/6 - 0.1, 2/3});
    \draw (d1) -- (d2);
    
    \coordinate (d3) at ({8+2/6 + 0.2, 4/3});
    \draw (d2) -- (d3);
    
    \coordinate (d4) at ({8+2/3, 4/3 + 0.2});
    \draw (d3) -- (d4);
    
    \coordinate (d5) at ({8+4/3, 4/3 + 0.4});
    \draw (d4) -- (d5);
    
    \coordinate (d6) at ({8+4/3 + 0.2, 4/3});
    \draw (d5) -- (d6);
    
    \coordinate (d7) at ({8+4/3 + 0.3, 2/3});
    \draw (d6) -- (d7);
    
    \coordinate (d8) at ({8+4/3, 2/3 - 0.2});
    \draw (d7) -- (d8);
    
    \coordinate (d9) at ({8+2/3, 2/3 - 0.4});
    \draw (d8) -- (d9);
    
    \coordinate (d10) at ({8+1, 1});
    \coordinate (d11) at ({8+4/3, 1.2});
    \draw (d10) -- (d11);
    
    \coordinate (m1) at ({10 - 2/6, 1});
    \coordinate (m2) at ({10 - 2/6 + 0.05, 2/3});
    \draw[densely dotted] (m1) -- (m2);
    
    \coordinate (m3) at ({8 + 4/3, 2/3 - 0.2 - 0.05});
    \draw[densely dotted] (m2) -- (m3);
    
    \coordinate (m4) at ({8 + 2/3, 2/3 - 0.2 - 0.05 - 0.2});
    \draw[densely dotted] (m3) -- (m4);
    
    \coordinate (m5) at ({8 + 2/3 - 1/6, 2/3});
    \draw[densely dotted] (m4) -- (m5);
    
    \coordinate (m6) at ({8+2/3, 2/3 + 0.5});
    \draw[densely dotted] (m5) -- (m6);
    
    \coordinate (m7) at ({8+2/3 + 0.3, 4/3});
    \draw[densely dotted] (m6) -- (m7);
    
    \coordinate (m8) at ({8+4/3, 4/3 + 0.15});
    \draw[densely dotted] (m7) -- (m8);
    
    \coordinate (m9) at ({8+4/3 +0.1, 4/3});
    \draw[densely dotted] (m8) -- (m9);
    
    \coordinate (m10) at ({8+2/6, 5/3});
    \coordinate (m11) at ({8+2/6 + 0.2, 2});
    \draw[densely dotted] (m10) -- (m11);
    \end{tikzpicture}
    \caption{Stacked naive trackings for $\mathcal{P}_0$ tracking.}
    \label{fig:stacked_naive_trackings}
\end{figure}
In Fig. \ref{fig:stacked_naive_trackings}, the naive tracking stacking is illustrated. Since the initial cell sampled is random, even if all trajectories are parallel within a cell, the cells crossed by the trajectory might differ. This scenario is depicted in Fig. \ref{fig:stacked_naive_trackings}. 

In the first case, the initial cell sampled is the leftmost bottom cell, and the tracking creates a loop. The center cell is tracked and reaches a calculable cell belonging to the loop. The tracking is therefore completed as all cells are calculable. 

In the second case, the initial cell sampled is the middle right cell. Since a different cell is chosen, a different loop is generated. The upper left cell is not initially calculable, but it is tracked and subsequently becomes calculable.

In Fig. \ref{fig:stacked_naive_trackings}, the two trackings are then stacked to compute the concentration within the cell. The resulting volume-averaged concentration is the weighted time-of-flight average of the line-averaged concentrations from each naive tracking.

\subsubsection{The submesh method}\label{subsec:submesh}
In the submesh approach, each cell is subdivided into a set of subcells. The values of the original velocity field, possibly on a coarser mesh, are duplicated within the subcells. A single tracking is then performed on the entire mesh. The DNPs concentration within a cell is obtained by computing the weighted average of the concentrations within the subcells. The weights are given by Eq. \eqref{eq:vol_average}, with the \(T_i^{(k)}\) replaced by the time-of-flights within the subcells. 

The main advantage of the submesh method is that it forces the tracks to pass through each part of the cell. However, the submesh method is expected to be more computationally expensive than the naive tracking stacking because it requires tracking a larger number of pathlines.

Submesh trackings can also be stacked in the same manner as naive trackings, as shown in Fig. \ref{fig:stacked_naive_trackings}.
\section{Coupling with the neutron transport equation}\label{sec:coupling}
The DNPs balance equations are solved together with the multigroup neutron transport equation. In turn, this balance equation is solved using the discrete ordinates or $S_\mathrm{N}$ method on a structured mesh \cite{lewis1984computational}, as described in Sec. \ref{subsec:neutron_balance}. Nuclear data is taken from the benchmark properties \cite{tiberga2020results}. The velocity field is obtained using TrioCFD developed at the French Alternative Energies and Atomic Energy Commission (CEA) \cite{angeli2015overview}.
\subsection{Multigroup Neutron Balance Equation}\label{subsec:neutron_balance}
In the steady state, the multigroup neutron transport equation is a balance equation between streaming, absorption, scattering and production,
\begin{equation}
    \vb*{\Omega}\cdot\grad{\psi_g} + \Sigma_t^g \psi_g = \mathcal{S}_g\psi + \frac{1}{\keff}\mathcal{F}_p \psi + \mathcal{F}_d \psi,
    \label{eq:neutron_transport}
\end{equation}
where $\psi_g$ is the angular neutron flux in group $g$, $\Sigma_t^g$ is the total macroscopic cross-section. $\mathcal{S}_g\psi$ is the scattering source term,
\begin{equation*}
    \mathcal{S}_g\psi = \sum_{g^\prime=1}^{G}\sum_{\ell=0}^{L}\frac{2\ell + 1}{4\pi}\Sigma_{s\ell}^{g^\prime\to g}\sum_{m=-\ell}^{\ell}R_{\ell m}\qty(\vb*{\Omega}) \phi^{\ell m}_{g^\prime},
\end{equation*}
where the $\Sigma_{s\ell}$ are the $\ell$-th order scattering macroscopic cross-sections, developed on a Legendre polynomial basis \cite{lewis1984computational}. The $R_{\ell m}$ functions are the real spherical harmonics, and are Schmidt semi-normalized \cite{wieczorek2018shtools}. $\phi^{\ell m}_{g^\prime}$ are the $(\ell, m)$ angular flux moments in the energy group $g^\prime$, defined as,
\begin{equation}
    \phi^{\ell m}_{g^\prime}\qty(\vb{r}) = \int_{4\pi}\dd{\vb*{\Omega}}R_{\ell m}\qty(\vb*{\Omega})\psi_{g^\prime}\qty(\vb{r}, \vb*{\Omega}).
    \label{eq:angular_flux_moments}
\end{equation}
Then, the prompt fission term of Eq. \eqref{eq:neutron_transport} is given by,
\begin{equation*}
    \mathcal{F}_p \psi = \frac{(1 - \beta)\chi_p^g}{4\pi}\sum_{g^\prime=1}^{G}\nu^{g^\prime}\Sigma_f^{g^\prime}\int_{4\pi}\dd{\vb*{\Omega}^\prime}\psi^{g^\prime} = \frac{(1 - \beta)\chi_p^g}{4\pi}\sum_{g^\prime=1}^{G}\nu^{g^\prime}\Sigma_f^{g^\prime}\phi_{g^\prime}^{00},
\end{equation*}
with $\chi^g$ is the fission spectrum, $\nu^{g^\prime}$ is the average number of neutrons produced per fission in group $g^\prime$, $\Sigma_f^{g^\prime}$ is the fission macroscopic cross-section, $\phi_{g^\prime}^{00}$ is the scalar neutron flux in group $g^\prime$ and $\beta = \sum_j \beta_j$. The delayed fission source, $\mathcal{F}_d\psi$ and the delayed neutron precursors source \(S_j\) are given by,
\begin{equation}
    \mathcal{F}_d\psi = \frac{\chi_d^g}{4\pi} \sum_j \lambda_j C_j \qq{and} S_j = \frac{\beta_j}{\keff} \sum_{g=1}^{G}\nu^g\Sigma_f^g\phi_g^{00},
    \label{eq:delayed_fission_source}
\end{equation}
where $\chi_d^g$ is the delayed neutron spectrum. The neutron flux does not directly appear in \(\mathcal{F}_d\), Eq. \eqref{eq:delayed_fission_source}, but concentrations of DNPs are linear in flux through the source term of Eq. \eqref{eq:precursors_advection}. In Eqs. \eqref{eq:neutron_transport} and \eqref{eq:delayed_fission_source}, the effective multiplication factor $\keff$ is introduced as $\nu^g \to \nu^g / \keff$ and is playing the role of eigenvalue for the homogeneous problem. In addition, a void condition is applied on the angular neutron flux at the boundaries of the domain,
\begin{equation}
    \psi^g\qty(\vb{r}, \vb*{\Omega}) = 0 \qq{for} \vb{r} \in \partial\Omega \qq{and} \vb*{\Omega}\cdot\vu{n} > 0.
    \label{eq:neutron_bc}
\end{equation}
\subsection{Angular discretization}\label{subsec:angular_discretization}
The discrete directions on which Eq. \eqref{eq:neutron_transport} is solved are selected in the half-hemisphere due to the symmetry of the problem with respect to the $x-y$ plane. 

The azimuthal and polar angles, respectively $\varphi$ and $\mu=\cos\theta$ of the direction vector \(\vb*{\Omega} = \qty(\sqrt{1-\mu^2}\cos\varphi, \sqrt{1-\mu^2}\sin\varphi)\) are discretized using the Gauss-Legendre and Gauss-Tchebychev quadratures. The flux moments, Eq. \eqref{eq:angular_flux_moments} are calculated as a weighted sum of the discrete angular fluxes \(\phi^{\ell m}_g = \sum_{n=1}^{N_{\mu}}\sum_{k=1}^{N_{\varphi}} w_n w_k  R_{\ell m}\qty(\vb*{\Omega}_{nk})\psi_g\qty(\vb{r}, \vb*{\Omega}_{nk})\) where $w_n$ and $w_k$ are respectively the weights of the Gauss-Legendre and Tchebychev-Gauss quadratures, and $\vb*{\Omega}_{nk}$ is the direction vector associated to the $n$-th polar angle and the $k$-th azimuthal angle \cite{abramowitz1968handbook}. The number of polar and azimuthal angles are respectively $N_{\theta}$ and $N_{\varphi}$.
\subsection{Space discretization}\label{subsec:space_discretization}
The space discretization of Eq. \eqref{eq:neutron_transport} is performed using a diamond difference scheme \cite{lewis1984computational}. In the following equations of this section, the energy group index is omitted. The angular flux is integrated over the control volume $V_k$,
\begin{equation*}
    \psi_{i,j} = \frac{1}{\Delta x \Delta y}\int_{x_{i-1/2}}^{x_{i+1/2}}\int_{y_{j-1/2}}^{y_{j+1/2}}\dd{x}\dd{y}\psi\qty(x,y).
    \label{eq:angular_flux_discretization}
\end{equation*}
Cross-sections are assumed to be constant within the cell. The diamond approximation is used to relate the numerical fluxes at the cell interfaces to the mean angular flux values over a cell \cite{lewis1984computational},
\begin{align}
    \psi_{i+1/2,j} &= 2\psi_{i,j} - \psi_{i-1/2,j}, \qq{if}\vb*{\Omega}\cdot\vu{x} > 0,\label{eq:sweep_left_to_right} \\
    \psi_{i-1/2,j} &= 2\psi_{i,j} - \psi_{i+1/2,j}, \qq{if}\vb*{\Omega}\cdot\vu{x} < 0,\label{eq:sweep_right_to_left} \\
    \psi_{i,j+1/2} &= 2\psi_{i,j} - \psi_{i,j-1/2}, \qq{if}\vb*{\Omega}\cdot\vu{y} > 0,\label{eq:sweep_bottom_to_top} \\
    \psi_{i,j-1/2} &= 2\psi_{i,j} - \psi_{i,j+1/2}, \qq{if}\vb*{\Omega}\cdot\vu{y} < 0.\label{eq:sweep_top_to_bottom}
\end{align}
The boundary condition Eq. \eqref{eq:neutron_bc} is prescribed by setting a zero upwind numerical flux on the boundaries. In Eqs. \eqref{eq:sweep_left_to_right} to \eqref{eq:sweep_top_to_bottom}, $\vu{x}$ and $\vu{y}$ are the unit vectors along the $x$ and $y$ axis, respectively.
\subsection{Iterative scheme}\label{subsec:iterative_scheme}
The steady state neutronic problem defined by Eqs. \eqref{eq:neutron_transport}, \eqref{eq:neutron_bc} and \eqref{eq:precursors_advection} is solved using the power iteration method \cite{lewis1984computational}. At the beginning of the calculation, a MOC pre-calculation is performed on the given velocity field to store the \textit{tracking data} needed to solve Eq. \eqref{eq:precursors_advection} according to the numerical method defined in Sec. \ref{sec:calculation_strategy}. The convergence criterion on the flux moments is set to,
\begin{equation}
    \frac{\norm{\phi_{\ell m}^{(n+1)} - \phi_{\ell m}^{(n)}}}{\norm{\phi_{\ell m}^{(n)}}} < \num{1e-5} \quad \forall \qty(\ell, m).
    \label{eq:flux_moments_convergence}
\end{equation}
The multigroup scattering source follows the convergence criterion of Eq. \eqref{eq:flux_moments_convergence} for all energy groups. The fission source is treated with the power iteration method, and is considered converged when,
\begin{equation}
    \frac{\abs{\keff^{(n+1)} - \keff^{(n)}}}{\keff^{(n)}} < \num{1e-7},
    \label{eq:keff_convergence}
\end{equation}
with an updated multiplication factor given by,
\begin{equation*}
    \keff^{(n+1)} = \sqrt{\frac{\int\dd{V} \qty[\qty(1-\beta)\sum_g \nu^g \Sigma_{f}^g \phi_{g}^{(n+1)} + \sum_j \lambda_j C_j^{(n+1)}]^2}{\int\dd{V} \qty[(1-\beta)\sum_g \nu^g \Sigma_{f}^g \phi_{g}^{(n)} + \sum_j \lambda_j C_j^{(n)}]^2}},
\end{equation*}
and the $\nu$-fission source calculated at the $(n+1)$-iteration is divided by $\keff^{(n+1)}$. The delayed source is updated at each iteration according to the method presented in Sec. \ref{sec:calculation_strategy}.
\section{Results}\label{sec:results}
This section features the results of the neutronic phases of the benchmark, specifically steps 0.2 and 1.1, with no power coupling. The first step, 0.1, is presented in Sec. \ref{subsec:static_fuel} while step 1.1 is presented in Sec. \ref{subsec:step_1.1}.
\subsection{Neutronic with static fuel}\label{subsec:static_fuel}
The first neutron physics step of the benchmark is the static fuel. The fuel velocity is set to zero, and Eq. \eqref{eq:precursors_advection} is analytically inverted and used in Eq. \eqref{eq:neutron_transport}. This defines an effective neutron spectrum, $\hat{\chi}^g = (1-\beta)\chi_p^g + \beta\chi_d^g$. The anisotropy order is set to $L=3$ and the number of energy groups is set to $G=6$, according to the nuclear data of the benchmark \cite{tiberga2020results}. 
The results are consistent with reactivities calculated using other deterministic codes \cite{tiberga2020results,greiner2023new,griffinCNRS2024}. Since the difference in reactivities computed with $(8,6)$ and $(8,8)$ falls below the convergence threshold (Eq. \eqref{eq:keff_convergence}), the angular discretization is set to $(8,6)$. The mesh convergence of the neutron transport equation was tested by varying the number of cells and computing the static reactivity. For a \(50 \times 50\) mesh, the reactivity difference compared to the \(100 \times 100\) mesh is \SI{4}{\pcm}. Between the \(100 \times 100\) and \(200 \times 200\) meshes, the reactivity difference is less than \SI{0.2}{\pcm}. Thus, mesh convergence is considered achieved. Finally, the reactivity difference between the \(200 \times 200\) and \(400 \times 400\) meshes is below the convergence threshold defined in Eq. \eqref{eq:keff_convergence}.
\subsection{Circulating fuel}\label{subsec:step_1.1}
In this subsection, the numerical method developed in Sec. \ref{sec:general_precursors_calculation_scheme} is coupled with the neutron transport equation described in Sec. \ref{sec:coupling}. In Sec. \ref{subsubsec:zero_order_interpolation} (respectively Sec. \ref{subsubsec:first_order_interpolation}), \(\mathcal{P}_0\) tracking (respectively \(\mathcal{P}_1\) tracking) is used. For both tracking methods, the reactivity difference between the static and circulating fuel is computed and compared to the values provided by other benchmark participants. The total delayed source along \(\mathrm{A}\mathrm{A}^\prime\) and \(\mathrm{B}\mathrm{B}^\prime\) (see Fig. \ref{fig:lid_cavity}) is also calculated as a function of the number of trackings. The two numerical schemes for evaluating the concentration within a cell, presented in Sec. \ref{sec:general_precursors_calculation_scheme}, are tested. A map of time-of-flights is shown in Sec. \ref{par:map_of_time_of_flights}, highlighting the recirculation zones of the fuel.
\subsubsection{$\mathcal{P}_0$ tracking}\label{subsubsec:zero_order_interpolation}
\begin{figure}[h!]
\centering
\includegraphics[width=0.8\textwidth]{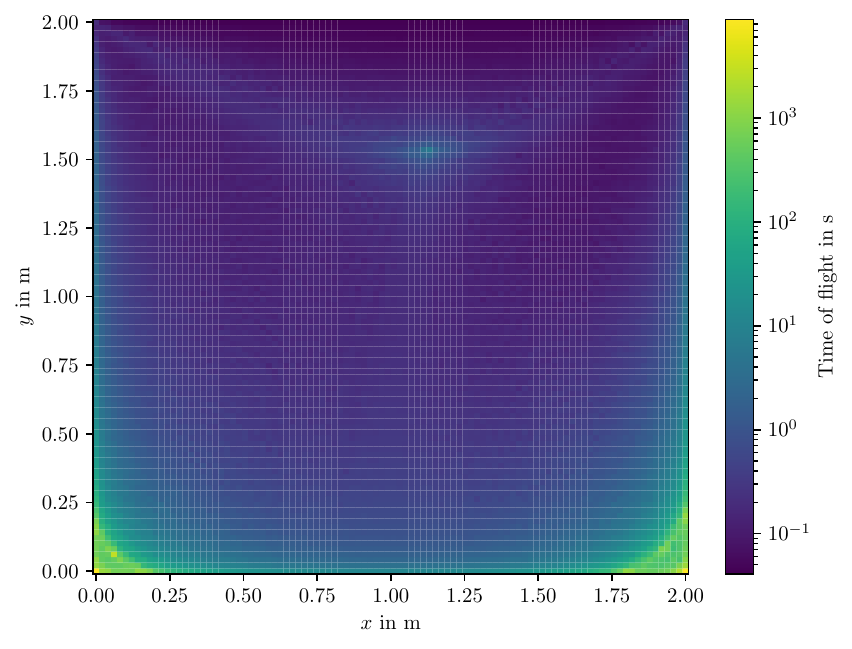}
\caption{Map of time-of-flights.}
\label{fig:map_of_time_of_flights}
\end{figure}

\begin{figure}[h!]
	\centering
	\begin{subfigure}[b]{0.45\textwidth}
		\includegraphics[width=\textwidth]{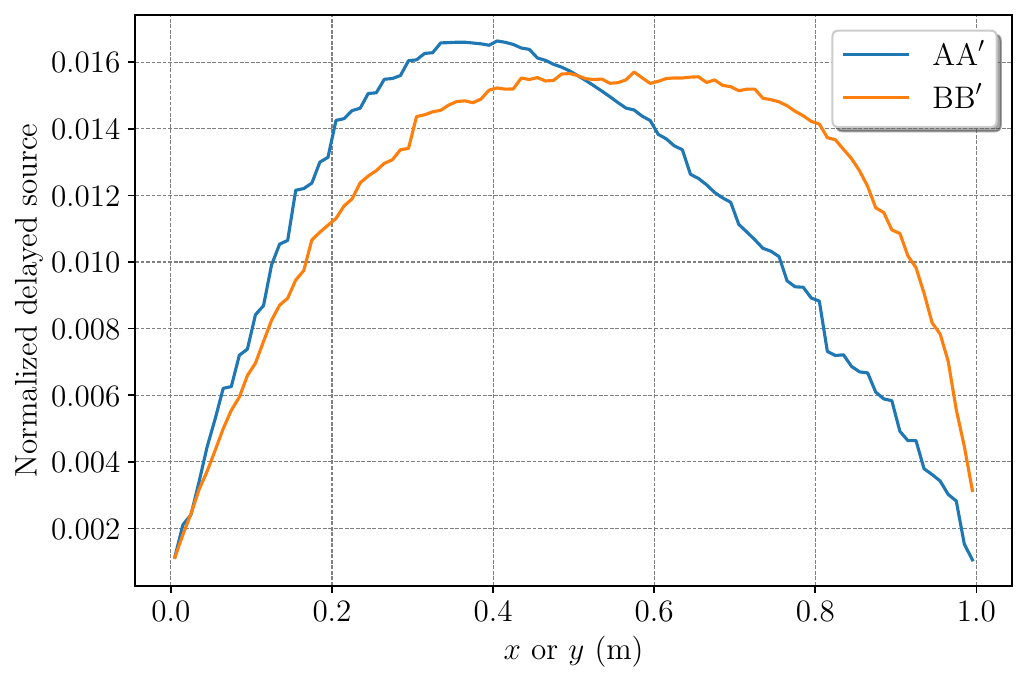}
		\caption{Normalized delayed source computed with one tracking.}
		\label{fig:delayed_source_1_trackings_P0}
	\end{subfigure}
	\hfill
	\begin{subfigure}[b]{0.45\textwidth}
		\includegraphics[width=\textwidth]{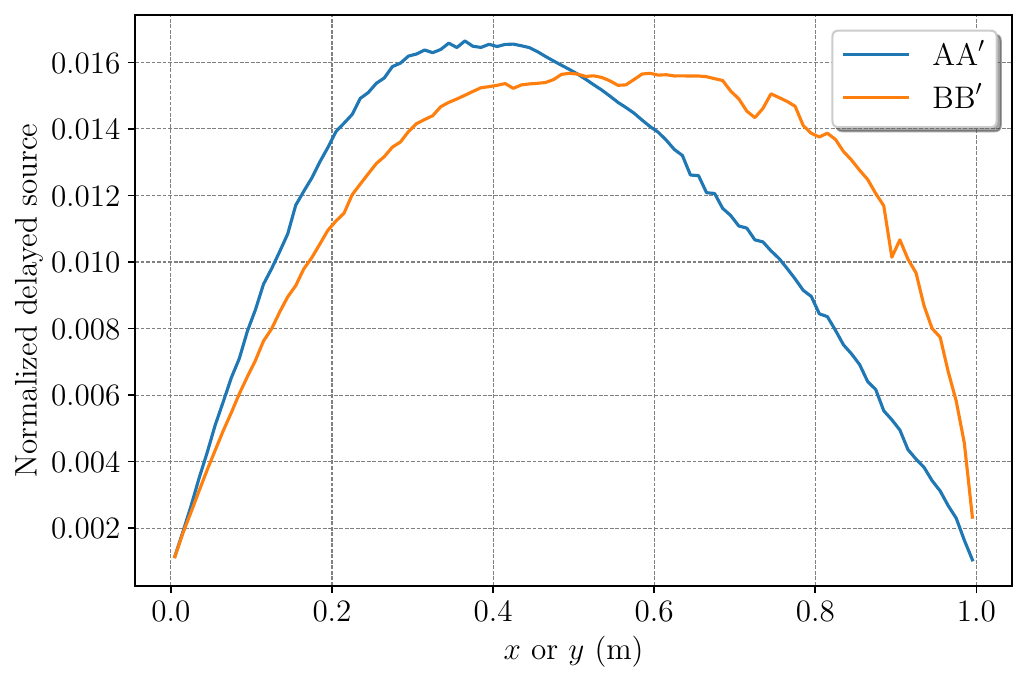}
		\caption{Normalized delayed source computed with twenty trackings.}
		\label{fig:delayed_source_20_trackings_P0}
	\end{subfigure}
	\caption{Total delayed source along $\mathrm{A}\mathrm{A}^\prime$ and $\mathrm{B}\mathrm{B}^\prime$ for different number of naive $\mathcal{P}_0$ trackings stacked.}
	\label{fig:delayed_source_zero_order}
\end{figure}

\begin{figure}[h!]
	\centering
	\begin{subfigure}[b]{0.45\textwidth}
		\includegraphics[width=\textwidth]{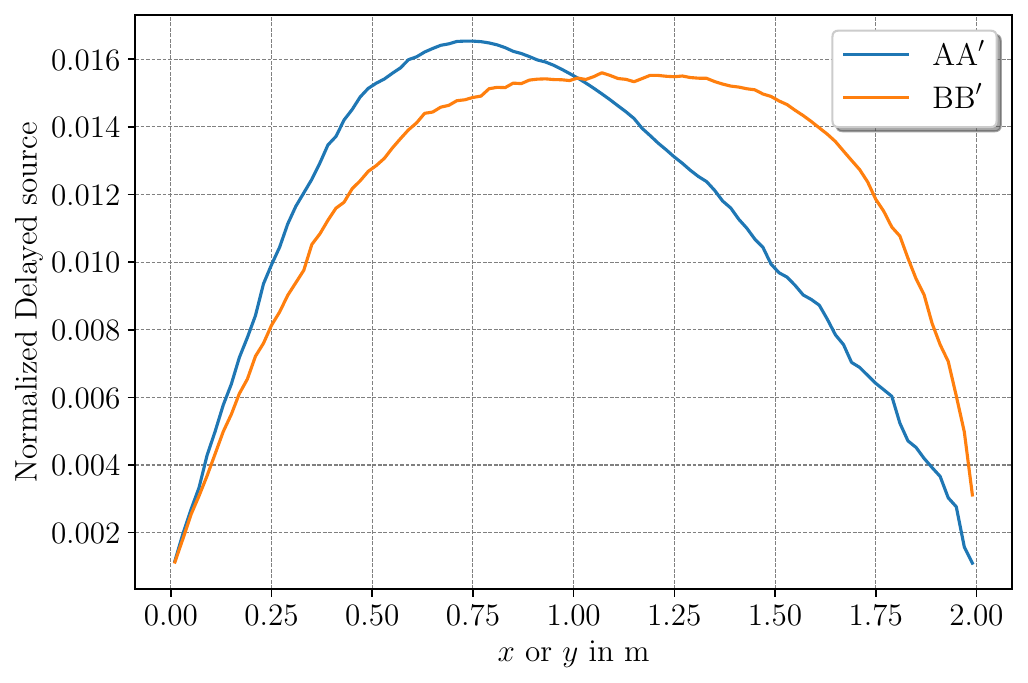}
		\caption{Normalized delayed source obtained with one $\mathcal{P}_0$ submesh tracking.}
		\label{fig:delayed_source_false_submesh}
	\end{subfigure}
	\hfill
	\begin{subfigure}[b]{0.45\textwidth}
		\includegraphics[width=\textwidth]{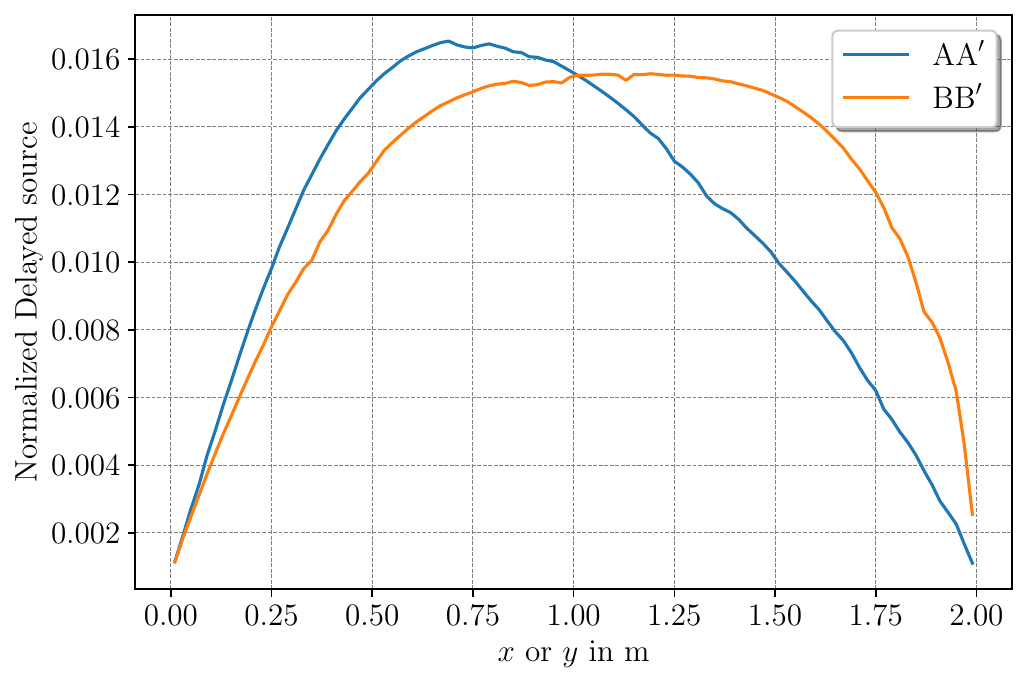}
		\caption{Normalized delayed source obtained with twenty $\mathcal{P}_0$ submesh tracking.}
		\label{fig:delayed_source_true_submesh}
	\end{subfigure}
	\caption{Total delayed source along $\mathrm{A}\mathrm{A}^\prime$ and $\mathrm{B}\mathrm{B}^\prime$ for a different number of $\mathcal{P}_1$ submesh trackings.}
	\label{fig:delayed_source_zero_order_submesh}
\end{figure}

\begin{figure}[h!]
	\centering
	\begin{subfigure}[b]{0.45\textwidth}
		\includegraphics[width=\textwidth]{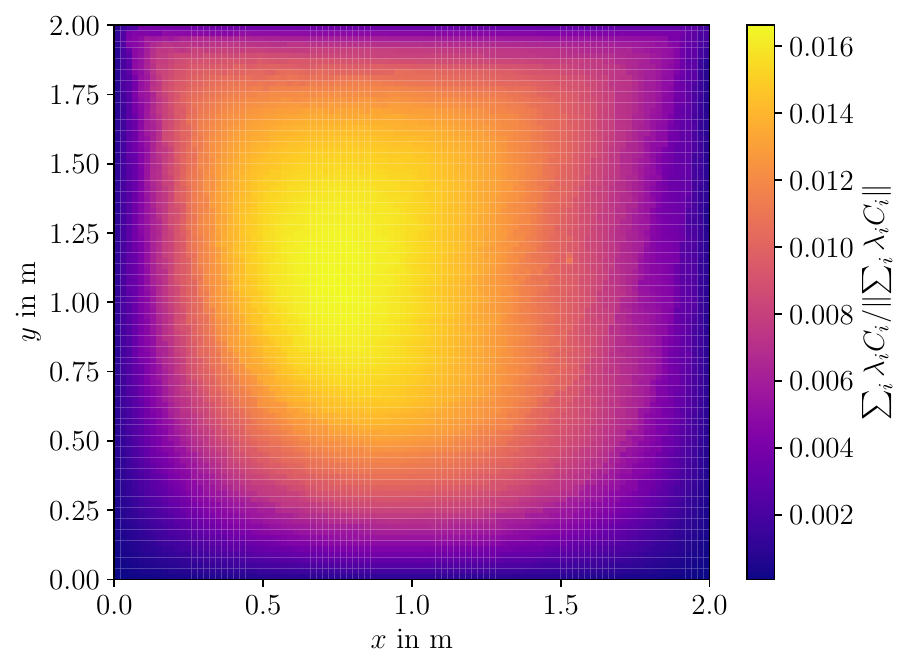}
		\caption{Normalized delayed source computed using four trackings.}
		\label{fig:2D_delayed_source_four_P0_trackings}
	\end{subfigure}
	\hfill
	\begin{subfigure}[b]{0.45\textwidth}
		\includegraphics[width=\textwidth]{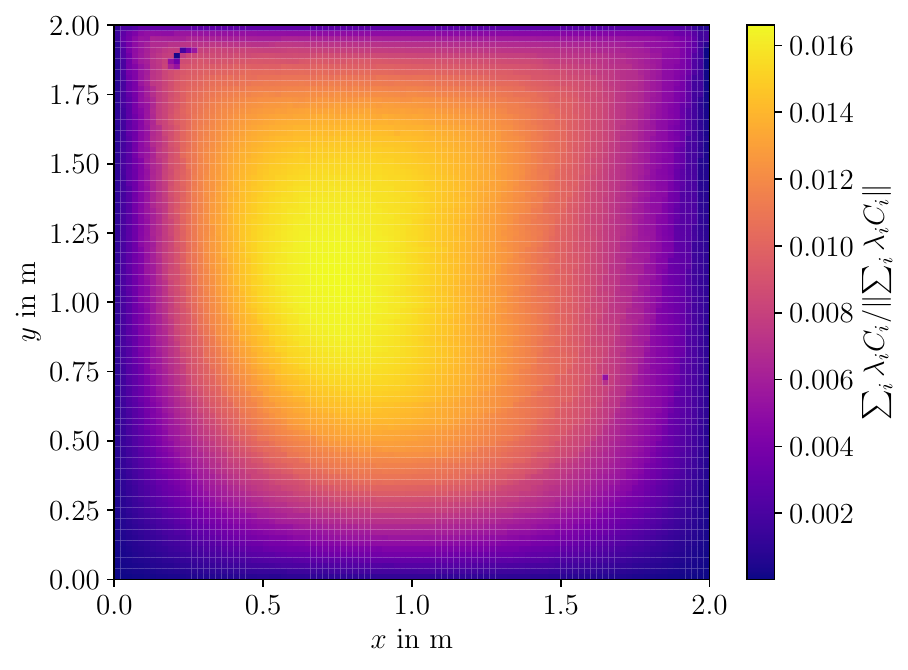}
		\caption{Normalized delayed source using a submesh approach with a duplicated velocity field.}
		\label{fig:2D_delayed_source_submesh_trackings}
	\end{subfigure}
	\caption{Total delayed source in the cavity for two different MOC approach.}
	\label{fig:delayed_source_2D}
\end{figure}

\begin{figure}[h!]
\centering
\includegraphics[width=0.8\textwidth]{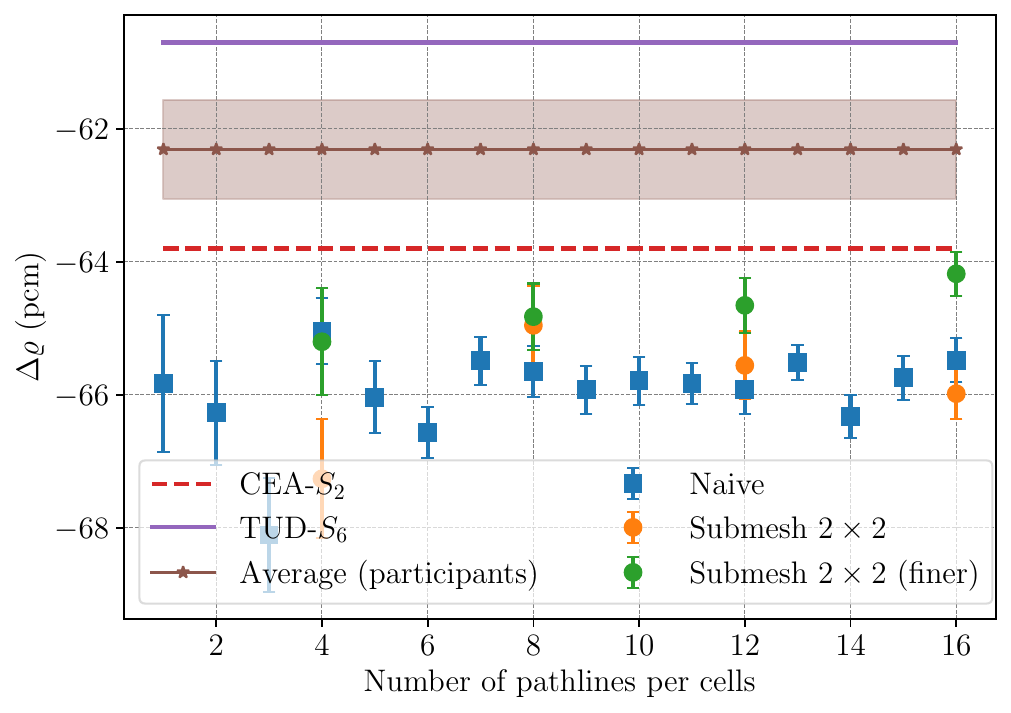}
\caption{Reactivity difference between the static and circulating fuel as a function of the number of $\mathcal{P}_0$ pathlines per cell, with lower and upper bounds of the other deterministic codes, the average of the other deterministic codes, and the associated standard deviation.}
\label{fig:reactivity_difference_zero_order}
\end{figure}

\paragraph{Map of time-of-flights}\label{par:map_of_time_of_flights}
The map of time-of-flights is shown in Fig. \ref{fig:map_of_time_of_flights}, obtained by averaging 50 trackings.
Fig. \ref{fig:map_of_time_of_flights} shows that time-flights span across several orders of magnitude between the different regions of the cavity. In the two bottom corner of the geometry, stagnation points are defined by very large time-of-flight per cell. On the contrary, cells located closer to the lid exhibits times-of-flight orders of magnitudes lower than in the stagnation points. This is because advection is stronger near the top of the cavity. A stagnation point also appears near the center of the cavity, which corresponds to the center of the vortex.
\paragraph{Naive tracking stacking}
Using multiple trackings, the total delayed source along $\mathrm{A}\mathrm{A}^\prime$ and $\mathrm{B}\mathrm{B}^\prime$ are also computed and shown in Fig. \ref{fig:delayed_source_zero_order}.
In Fig. \ref{fig:delayed_source_zero_order}, the total delayed source is computed with one and twenty trackings. The total delayed source calculated with one tracking is noisier than the one calculated with twenty trackings as an average of the line-averaged concentrations is performed through Eq. \eqref{eq:vol_average}. This was expected because stacking multiple trackings allows cumulating line-averaged values of DNPs concentrations within the same cell, which are then averaged with respect to the time-of-flight to obtain the volume-averaged concentration.
\paragraph{Submesh method}
One possible way to smoothen the concentrations is to use a submesh. Each cell is broken into $2\times 2$ subcells. The tracking is performed on the submesh, and the concentration is calculated in the mesh cell using a weighted sum of the line-averaged concentrations. This requires to represent the velocity field on the submesh. Therefore, the velocity vector of the original cell, which is constant along $\vu{x}$ and $\vu{y}$ is duplicated to the subcells. The total normalized delayed source is computed and shown in Fig. \ref{fig:delayed_source_zero_order_submesh} for one and twenty trackings on the submesh.
In Fig. \ref{fig:delayed_source_zero_order_submesh}, it can be seen that the total delayed source is smoother when computed with a larger number of submesh tracking. This is due to a second layer of averaging that averages the concentration obtained between the different trackings. The total delayed source within the cavity can also be represented for both the submesh method with a duplicated field and a naive stacking of four trackings, as shown in Fig. \ref{fig:delayed_source_2D}.
Both Fig. \ref{fig:2D_delayed_source_four_P0_trackings} and Fig. \ref{fig:2D_delayed_source_submesh_trackings} exhibit very similar results. The DNPs are advected by the flow and decay on the left-hand side of the cavity whereas with $\vb{u}=\vb{0}$ the delayed source would exhibit the same shape as the neutron flux.
\paragraph{Reactivity difference for both methods}
Using multiple trackings, the reactivity difference between the static and circulating fuel is computed and shown in Fig. \ref{fig:reactivity_difference_zero_order}. For each possible number of trackings, the reactivity calculation is repeated fifty times to estimate the uncertainty on the reactivity difference by estimating its standard deviation. The evolution of reactivity with the number of trackings is calculated for both the naive tracking stacking and the submesh method. 

Two submesh methods are tested, one with a duplicated velocity field from the coarser mesh, referred as ``submesh $2\times 2$'' and one with a velocity field coming from a CFD calculation on the submesh, referred as ``submesh $2\times 2$ finer''.
In Fig. \ref{fig:reactivity_difference_zero_order}, the reactivity difference appears to be fluctuating, independently of the number of trackings stacked, and independently of the MOC scheme used. These random variations of reactivity, above the convergence threshold of \SI{1e-1}{\pcm} of the power iteration, Eq. \eqref{eq:keff_convergence}, may be due to the random placement of the seed points, as described in Sec. \ref{sec:calculation_strategy}. The reactivities calculated also appear to be slightly off the reference values given by the other benchmark participants, Table \ref{tab:reactivity_difference} \cite{greiner2023new,tiberga2020results}.
\begin{table}[h!]
\centering
\begin{tabular}{cc}
\hline
Code & $\Delta \varrho$ (pcm) \\
\hline
\hline
CNRS-SP\textsubscript{1} & $-62.5$ \\
CNRS-SP\textsubscript{3} & $-62.6$ \\
PoliMi & $-62.0$ \\
PSI & $-63.0$ \\
TUD-S\textsubscript{2} & $-62.0$ \\
TUD-S\textsubscript{6} & $-60.7$ \\
CEA-diff & $-62.7$ \\
CEA-SP\textsubscript{1} & $-62.7$ \\
CEA-SP\textsubscript{3} & $-62.3$ \\
CEA-S\textsubscript{2} & $-63.8$ \\
CEA-S\textsubscript{4} & $-62.4$ \\
CEA-S\textsubscript{8} & $-62.3$ \\
Griffin-Pronghorn & $-61.4$ \\
DTU & \SI{-6\pm 1e1}{} \\
SEALION & \SI{-58 \pm 9}{} \\
Naive $\times 1$ & \SI{-66 \pm 2}{} \\
Submesh ($2\times 2$) & \SI{-64 \pm 1}{} \\
Submesh ($3\times 3$) & \SI{-66 \pm 1}{} \\
Naive ($4 \times 1$) & \SI{-66 \pm 1}{} \\
% Submesh ($2 \times \qty(2\times 2)$) & \SI{-66 \pm 1}{} \\
\hline
\end{tabular}
\caption{Reactivity difference between the static and circulating fuel.}
\label{tab:reactivity_difference}
\end{table}
The $\mathcal{P}_0$ submesh method does appear to perform better than the naive calculation strategy. The submesh method with a recalculated velocity field appears to better match the results of the benchmark for $16$ pathlines per cell. However, it should be reminded that using a submesh is costlier than the naive tracking stacking because the mesh contains more cells. To cover the whole domain with both methods, the $2\times 2$ submesh method needs around \SI{510\pm 2e1}{} pathlines whereas the naive method needs around \SI{130 \pm 2e1}{} pathlines. All the $\mathcal{P}_0$ are off the reference values of the benchmark by at least \SI{3}{\pcm}. This could also be a consequence of the $\mathcal{P}_0$ tracking having difficulties near the boundaries of the domain where pathlines must be pushed back into the geometry, Sec. \ref{subsec:first_order}.
\subsubsection{$\mathcal{P}_1$ tracking}\label{subsubsec:first_order_interpolation}
\begin{figure}[h!]
	\centering
	\begin{subfigure}[b]{0.45\textwidth}
		\includegraphics[width=\textwidth]{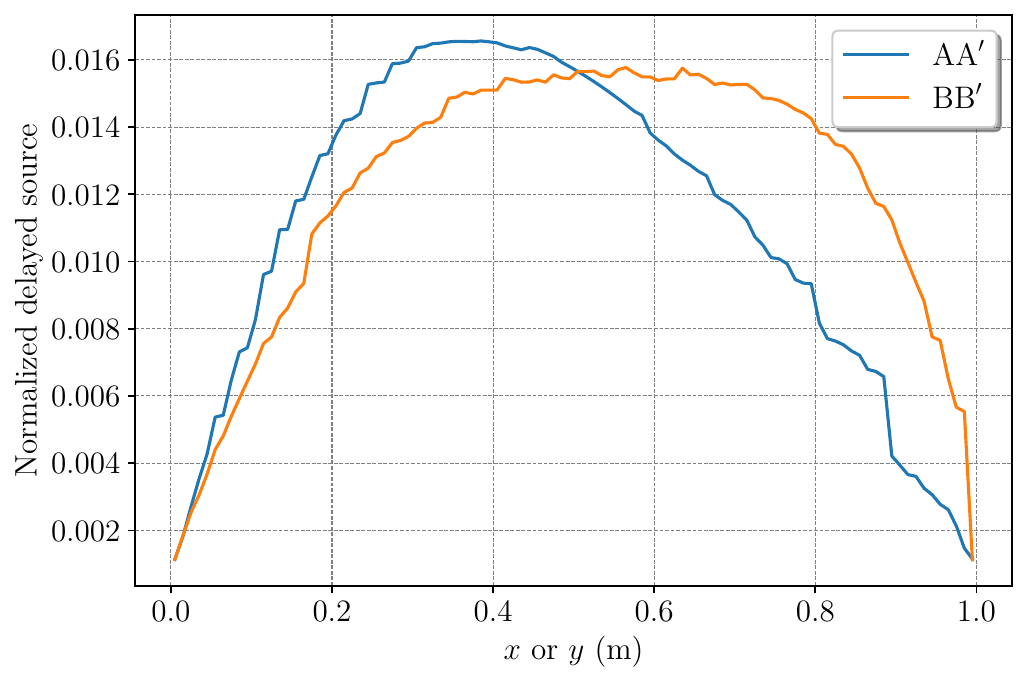}
		\caption{Normalized delayed source computed with one tracking.}
		\label{fig:delayed_source_1_trackings_P1}
	\end{subfigure}
	\hfill
	\begin{subfigure}[b]{0.45\textwidth}
		\includegraphics[width=\textwidth]{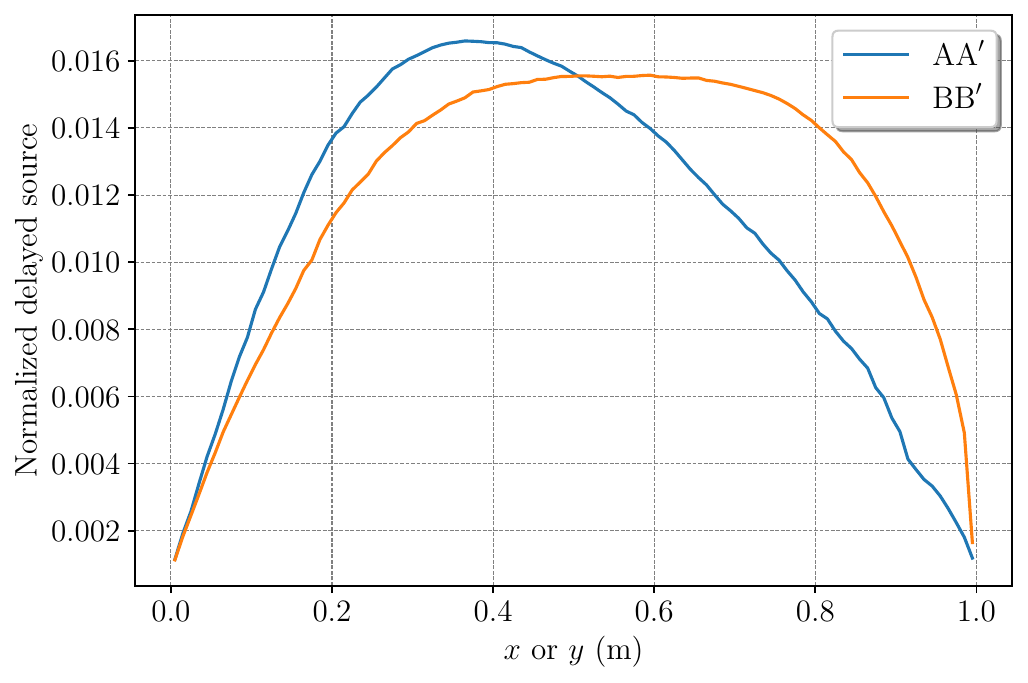}
		\caption{Normalized delayed source computed with twenty trackings.}
		\label{fig:delayed_source_20_trackings_P1}
	\end{subfigure}
	\caption{Total delayed source along $\mathrm{A}\mathrm{A}^\prime$ and $\mathrm{B}\mathrm{B}^\prime$ for different number of naive $\mathcal{P}_1$ trackings stacked.}
	\label{fig:delayed_source_first_order}
\end{figure}

\begin{figure}[h!]
	\centering
	\begin{subfigure}[b]{0.45\textwidth}
		\includegraphics[width=\textwidth]{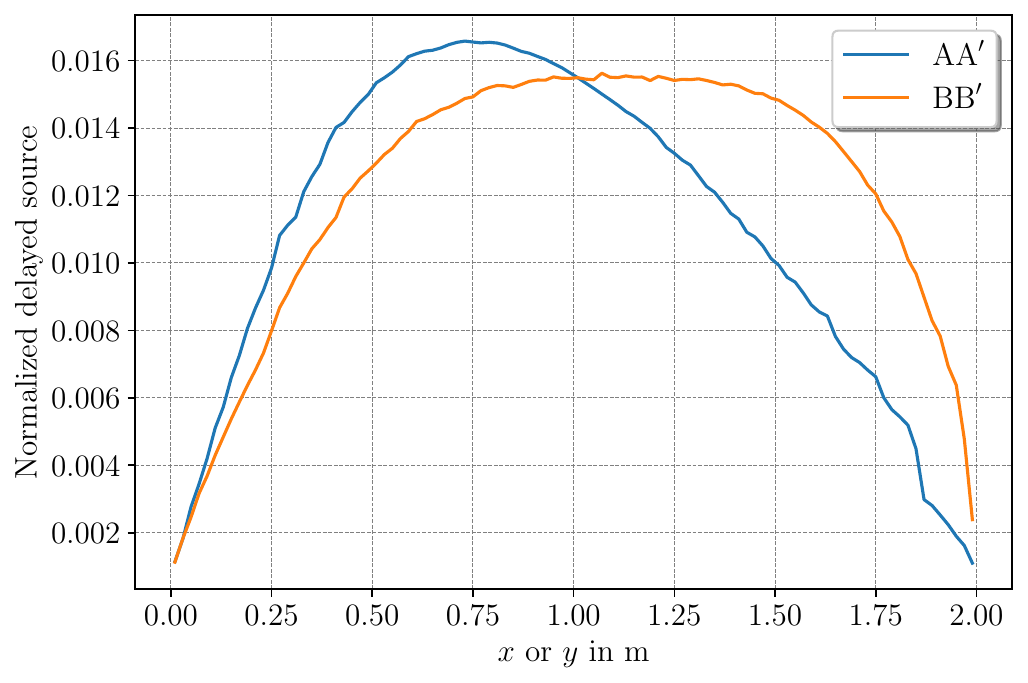}
		\caption{Normalized delayed source obtained with one $\mathcal{P}_1$ submesh tracking.}
		\label{fig:delayed_source_one_P1_submesh}
	\end{subfigure}
	\hfill
	\begin{subfigure}[b]{0.45\textwidth}
		\includegraphics[width=\textwidth]{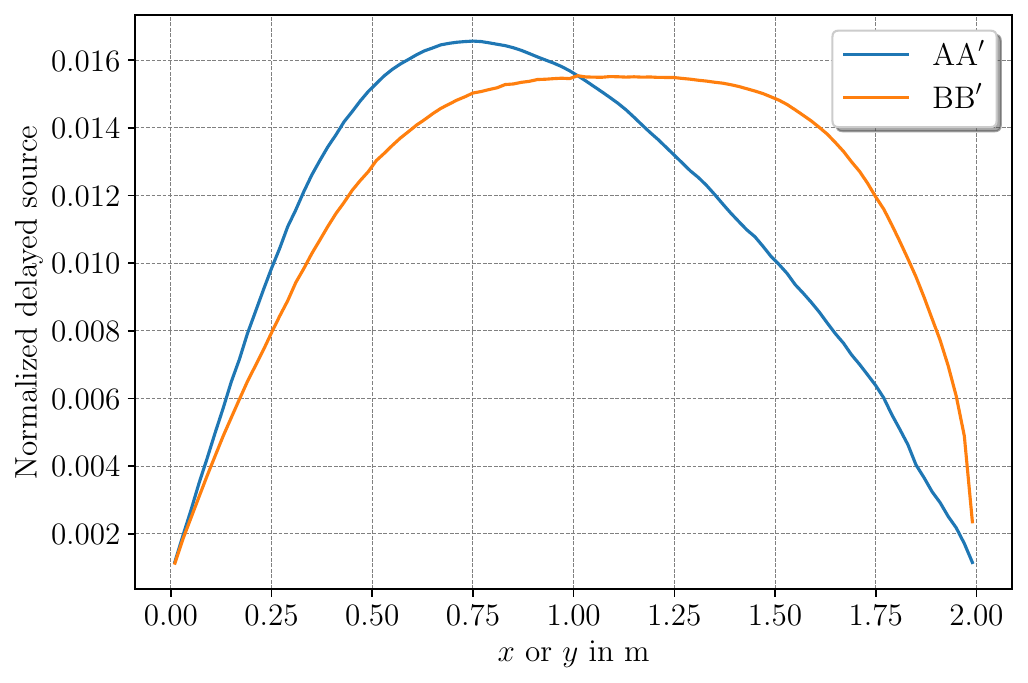}
		\caption{Normalized delayed source obtained with twenty $\mathcal{P}_1$ submesh tracking.}
		\label{fig:delayed_source_twenty_P1_submesh}
	\end{subfigure}
	\caption{Total delayed source along $\mathrm{A}\mathrm{A}^\prime$ and $\mathrm{B}\mathrm{B}^\prime$ for different submesh $\mathcal{P}_1$ trackings.}
	\label{fig:delayed_source_first_order_submesh}
\end{figure}

\begin{figure}[h!]
\centering
\includegraphics[width=0.8\textwidth]{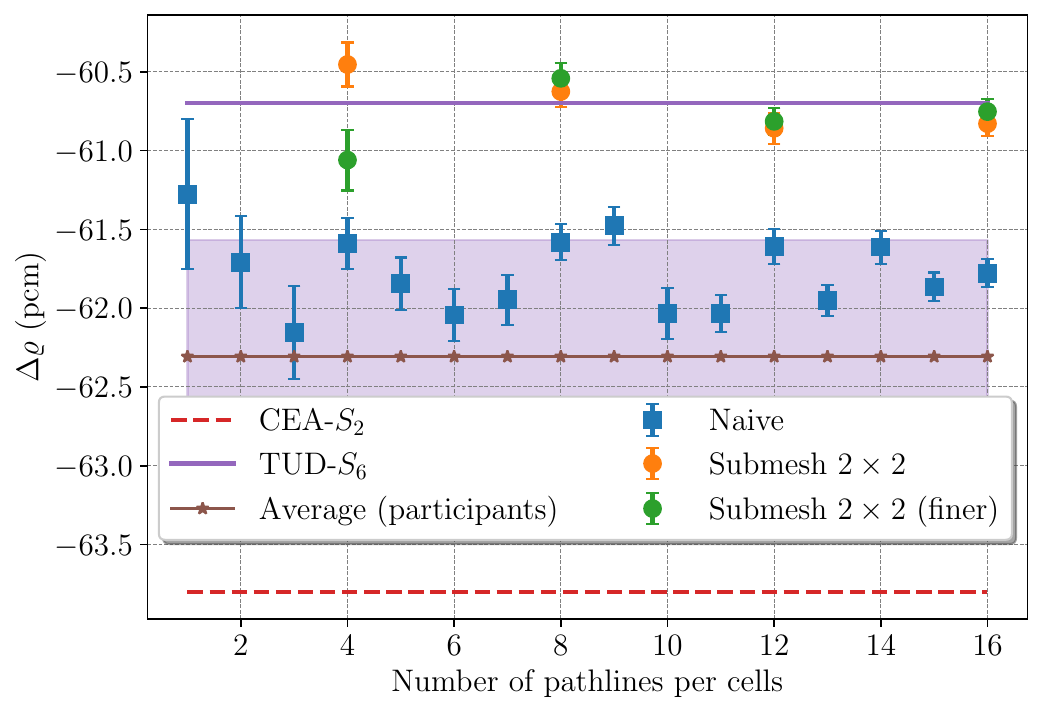}
\caption{Reactivity difference between the static and circulating fuel as a function of the number of $\mathcal{P}_1$ pathlines per cell, with lower and upper bounds of the other deterministic codes, the average of the other deterministic codes, and the associated standard deviation.}
\label{fig:reactivity_difference_first_order}
\end{figure}

In this section, the tracking method employs hyperbolic pathlines, as described in Sec. \ref{subsec:first_order}. As in Sec. \ref{subsubsec:zero_order_interpolation}, the delayed source along $\mathrm{A}\mathrm{A}^\prime$ and $\mathrm{B}\mathrm{B}^\prime$ are computed for different number of trackings, and different MOC approaches. The reactivity difference between the static and circulating fuel are calculated for both the naive tracking stacking and the submesh method.
\paragraph{Naive tracking stacking}
Using multiple $\mathcal{P}_1$ trackings, the total delayed source along $\mathrm{A}\mathrm{A}^\prime$ and $\mathrm{B}\mathrm{B}^\prime$ are computed and shown in Fig. \ref{fig:delayed_source_first_order}.
In Fig. \ref{fig:delayed_source_first_order}, the total delayed source is computed with one and twenty trackings. The total delayed source calculated with one tracking again appears noisier than the one calculated with twenty trackings. However, the total delayed source calculated with twenty tracking and the $\mathcal{P}_1$ tracking method is smoother than the one calculated with the $\mathcal{P}_0$ tracking method and twenty trackings, Fig. \ref{fig:delayed_source_20_trackings_P0}.
\paragraph{Submesh method}
The procedure can be repeated with a submesh, where the unknown inner velocity of the subcells faces are linearly interpolated. The total delayed source along $\mathrm{A}\mathrm{A}^\prime$ and $\mathrm{B}\mathrm{B}^\prime$ are computed and shown in Fig. \ref{fig:delayed_source_first_order_submesh}.
Again, the total delayed source computed with twenty trackings, Fig. \ref{fig:delayed_source_twenty_P1_submesh}, is smoother than the one obtained with one submesh tracking, Fig. \ref{fig:delayed_source_one_P1_submesh}.
\paragraph{Reactivity difference for both methods}
The reactivity differences between the static and circulating fuel are computed for both the naive tracking stacking and the submesh method. The results are shown in Fig. \ref{fig:reactivity_difference_first_order}. The statistical uncertainty is estimated by repeating fifty times the calculation with moving fuel.
In Fig. \ref{fig:reactivity_difference_first_order}, the reactivity difference appears less fluctuating than the reactivity difference obtained with $\mathcal{P}_0$ tracking. The reactivity differences obtained are also nearly all contained between the values of the other benchmark participants. For the same number of repeated calculation as in Sec. \ref{subsubsec:zero_order_interpolation}, the statistical uncertainty appears much smaller. For one tracking, the reactivity difference calculated is $\Delta \varrho = \SI{-61.3 \pm 0.5}{\pcm}$ which is close from the values of the other benchmark participants, Table \ref{tab:reactivity_difference}. For one submesh tracking, the reactivity difference obtained is $\Delta \rho = \SI{-60.3\pm 0.1}{\pcm}$, which is also consistent with the values obtained by the other participating codes, Table \ref{tab:reactivity_difference}. The mean reactivity difference values appear to be contained within the lower and upper bounds of the other participants of the benchmark. Both the submesh method with a coarser velocity field interpolated on the submesh, and the ``finer'' submesh method obtained similar results. The statistical uncertainty associated to one submesh tracking is lower than the one associated to one naive tracking. However, if both methods are compared on the basis of the number of pathlines per cell, their statistical uncertainty is similar. The values obtained with the $\mathcal{P}_1$ tracking method for four pathlines per cell are reminded in Table \ref{tab:P1_reactivities}.
\begin{table}[h!]
\centering
\begin{tabular}{cc}
\hline
Method & $\Delta \varrho$ (\SI{}{\pcm}) \\
\hline
\hline
Naive & \SI{-61\pm 0.2}{} \\
Submesh ($2\times 2$) & \SI{-60.3\pm 0.1}{} \\
Submesh (finer) & \SI{-60.9\pm 0.2}{} \\
\hline
\end{tabular}
\caption{Reactivity difference between the static and circulating fuel for the $\mathcal{P}_1$ tracking methods.}
\label{tab:P1_reactivities}
\end{table}

\subsubsection{Calculation time}\label{subsubsec:calculation_time}
For both tracking methods described in Sec. \ref{sec:pathlines_tracking}, the calculation time of the method described in Sec. \ref{sec:calculation_strategy} is given for the circulating fuel. The numerical methods described in this work are implemented in Python 3.11.8 using the NumPy 1.26.4 library. The neutron transport code as well as the DNPs transport code are run on a single core of an Intel Core i7-10610U CPU clocked at \SI{1.8}{\giga\hertz}, with \SI{32}{\giga\byte} of RAM. The calculation time is given in Table \ref{tab:calculation_time}, five calculations are performed to estimate the uncertainty on the calculation time. The neutron transport solver is initialized with a flat angular neutron flux.
\begin{table}[h!]
\centering
\begin{tabular}{ccc}
\hline
Step & Total time (\SI{}{\second}) ($\mathcal{P}_0$) & Total time (\SI{}{\second}) ($\mathcal{P}_1$) \\
\hline
\hline
$\qty(\times 1)$ tracking & \SI{0.74\pm 0.03}{} & \SI{2.6\pm 0.1}{} \\
$\qty(\times 1)$ submesh tracking & \SI{3.5\pm 0.2}{} & \SI{9.2\pm 0.3}{} \\
Power it. & \SI{201 \pm 2}{} & \SI{212\pm 2}{} \\
Power it. (submesh) & \SI{247 \pm 4}{} & \SI{227\pm 1}{} \\
DNPs calc. & \SI{3.47 \pm 0.06}{} & \SI{3.76\pm 0.04}{} \\
DNPs calc. (submesh) & \SI{14.9 \pm 0.2}{} & \SI{14.6\pm 0.2}{} \\
\hline
\end{tabular}
\caption{Total calculation time for each calculation step for a circulating fuel.}
\label{tab:calculation_time}
\end{table}
In Table \ref{tab:calculation_time}, for the $\mathcal{P}_0$ tracking method (i.e. straight lines within the cell mesh), the total tracking generation time is multiplied by four between the naive approach and the submesh method. This is consistent because the submesh needs nearly four times more tracks than the original mesh. The total calculation time of the delayed source is also multiplied by four. The total power iteration time is also slightly increased when using a submesh and a $\mathcal{P}_0$ tracking method.
The same trend is observed for the $\mathcal{P}_1$ tracking technique. When comparing both tracking methods for the same calculation step, the $\mathcal{P}_1$ tracking method appears to be more computationally expensive than the $\mathcal{P}_0$ tracking method. The tracking time is multiplied by a factor three, for roughly the same number of tracks generated (\SI{130 \pm 2e1}{} for the $\mathcal{P}_0$ versus \SI{1290 \pm 4}{} for the $\mathcal{P}_1$). The total time spent doing power iterations is also increased. However, independently of the tracking technique chosen, the total calculation time of the delayed source is nearly the same. The $\mathcal{P}_1$ tracking method produces reactivities with less statistical uncertainty than the $\mathcal{P}_0$ tracking method and more in range with the values obtained with other codes, see Sec. \ref{subsubsec:zero_order_interpolation} and \ref{subsubsec:first_order_interpolation}.
\section{Discussion \& Conclusion}\label{sec:conclusion}
The numerical method developed in this work aims at solving the DNPs balance equations of an incompressible liquid nuclear fuel at the steady state. The approach hinges upon the utilization of the Method of Characteristics (MoC), wherein pathlines are the characteristics of the DNPs advection balance equation. The advantage of such a method is that the \textit{tracking data} needed to compute the delayed source is stored during the MOC pre-calculation. This \textit{tracking data} is then used at each power iteration to evaluate the delayed neutron source term with the updated $\nu$-fission rate, therefore saving computational costs.

The pathline generation strategy presented here successfully produces \textit{tracking data} that covers the whole mesh. Different pathline trackings have been developed depending on the level of discretization of the given velocity vector field. Then, given some \textit{tracking data}, the DNPs concentration within a cell can be evaluated using the discretized solutions of the MOC. Two methods have been developed to evaluate the concentration, stacking \textit{tracking data}, referred to as ``naive tracking stacking'' throughout this work, and the submesh method that divides each cell into subcells. The advantage of the latter being that it forces pathlines to cross each part of the cell.

The method is tested on a MSRs benchmark case, the lid-driven cavity problem. The MOC DNPs strategy is coupled to an in-house discrete ordinate ($S_\mathrm{N}$) solver. This in-house solver compared well in the case of a static fuel with the other benchmark participants. The MOC strategy is then tested with a circulating fuel. The velocity field was obtained using TrioCFD.

The results obtained with the MOC strategy are consistent with the results obtained by the other participants of the benchmark. The reactivity difference between the static and circulating fuel is similar to other deterministic codes. The two tracking techniques ($\mathcal{P}_0$ and $\mathcal{P}_1$), the naive tracking stacking and the submesh methods were tested. The $\mathcal{P}_0$ tracking technique was found to off the reference values of the benchmark by at least \SI{3}{\pcm}, for either the naive tracking stacking or the submesh method. The $\mathcal{P}_1$ tracking technique gave reactivity differences nearly all contained within the lower and upper bounds of the other benchmark participants. The statistical uncertainty associated to the $\mathcal{P}_1$ tracking was also found to be lower than the one associated to the $\mathcal{P}_0$ tracking. The DNPs distributions obtained with the $\mathcal{P}_1$ tracking method were smoother than the ones obtained with the $\mathcal{P}_0$ tracking method, potentially due to a better pathline representation.

All the methods developed were also compared in terms of calculation time. The $\mathcal{P}_1$ tracking method was found to be more computationally expensive than the $\mathcal{P}_0$ tracking method. For both tracking techniques, the submesh method was more computationally expensive than one naive tracking. Compared to the total calculation time, the tracking generation time and the total delayed source calculation time were found to be one or two orders of magnitudes lower than the total power iteration time.

It remains to extend the numerical method to turbulent flows arising in MSRs, in which the diffusive term of the DNPs balance equation might become non-negligible with respect to the advection term. Strategies can be devised for steady state turbulent flows, where pathlines can be calculated from the mean velocity field (thus suitable for both RANS and LES calculations) and are currently under investigation. Treatment of time dependent problems remains to be explored. Finally, the key element of the method, the tracking algorithms, should also be extended to more complex meshes, such as unstructured 2D triangular meshes or 3D tetrahedral meshes.
\bibliographystyle{unsrt}
\bibliography{bibfile.bib}
\end{document}